\documentclass[prx,aps,letterpaper,groupedaddress,twocolumn,superscriptaddress,floatfix,nofootinbib]{revtex4-2}

\usepackage{mathptmx}
\usepackage[utf8]{inputenc}
\usepackage[T1]{fontenc}

\newcommand{\al}{\alpha}

\newcommand{\la}{\lambda}

\newcommand{\cC}{{\mathcal C}}

\newcommand{\ket}[1]{|#1\rangle}
\newcommand{\bra}[1]{\langle#1|}

\usepackage{amssymb,amsmath,amsfonts}
\usepackage{mathtools}
\usepackage{mathrsfs}
\usepackage{graphicx}
\usepackage{color}
\usepackage{bm}
\usepackage{xspace,soul}
\usepackage{siunitx}
\usepackage{dsfont}
\usepackage{braket}
\usepackage{booktabs,multirow}
\graphicspath{{figures/}}

\usepackage{comment}
\usepackage{float}
\usepackage[vmargin=1in, hmargin=0.9in]{geometry}
\usepackage{times}
\usepackage[normalem]{ulem}
\usepackage[breaklinks=true,colorlinks=true,urlcolor=blue,linkcolor=blue,citecolor=blue]{hyperref}
\usepackage{xr}
\usepackage{soul}
\usepackage[sort&compress]{natbib}

\makeatletter
\@ifundefined{textcolor}{}
{\definecolor{BLACK}{gray}{0}
 \definecolor{WHITE}{gray}{1}
 \definecolor{RED}{rgb}{1,0,0}
 \definecolor{GREEN}{rgb}{0,.4,0}
 \definecolor{BLUE}{rgb}{0,0,1}
 \definecolor{CYAN}{cmyk}{1,0,0,0}
 \definecolor{MAGENTA}{cmyk}{0,1,0,0}
 \definecolor{YELLOW}{cmyk}{0,.3,1,0}}
 \definecolor{light-gray}{gray}{0.90}
\makeatother

\usepackage[dvipsnames, table]{xcolor}

\begin{document}

\title{Towards experimental classical verification of quantum computation} 

\author{Roman Stricker}
\affiliation{Institut f\"ur Experimentalphysik, Universit\"at Innsbruck, Innsbruck, Austria}
\author{Jose Carrasco}
\affiliation{Institute for Theoretical Physics, University of Innsbruck, Innsbruck Austria}
\author{Martin Ringbauer}
\affiliation{Institut f\"ur Experimentalphysik, Universit\"at Innsbruck, Innsbruck, Austria}
\author{Lukas Postler}
\affiliation{Institut f\"ur Experimentalphysik, Universit\"at Innsbruck, Innsbruck, Austria}
\author{Michael Meth}
\affiliation{Institut f\"ur Experimentalphysik, Universit\"at Innsbruck, Innsbruck, Austria}
\author{Claire Edmunds}
\affiliation{Institut f\"ur Experimentalphysik, Universit\"at Innsbruck, Innsbruck, Austria}
\author{Philipp Schindler}
\affiliation{Institut f\"ur Experimentalphysik, Universit\"at Innsbruck, Innsbruck, Austria}
\author{Rainer Blatt}
\affiliation{Institut f\"ur Experimentalphysik, Universit\"at Innsbruck, Innsbruck, Austria}
\affiliation{Institute for Quantum Optics and Quantum Information, Austrian Academy of Sciences, Innsbruck, Austria}
\author{Peter Zoller}
\affiliation{Institute for Theoretical Physics, University of Innsbruck, Innsbruck Austria}
\affiliation{Institute for Quantum Optics and Quantum Information, Austrian Academy of Sciences, Innsbruck, Austria}
\author{Barbara Kraus}
\affiliation{Institute for Theoretical Physics, University of Innsbruck, Innsbruck Austria}
\author{Thomas Monz}
\affiliation{Institut f\"ur Experimentalphysik, Universit\"at Innsbruck, Innsbruck, Austria}
\affiliation{Alpine Quantum Technologies GmbH, Innsbruck, Austria}

\begin{abstract}
With today's quantum processors venturing into regimes beyond the capabilities of classical devices~\cite{Go19,wu2021strong,zhong2020quantum}, we face the challenge to verify that these devices perform as intended, even when we cannot check their results on classical computers \cite{eisert2020quantum,GhKa19}. In a recent breakthrough in computer science~\cite{Ma18,Bra20,Vidick2021}, a protocol was developed that allows the verification of the output of a computation performed by an untrusted quantum device based only on classical resources. Here, we follow these ideas, and demonstrate in a first, proof-of-principle experiment a verification protocol using only classical means on a small trapped-ion quantum processor.
We contrast this to verification protocols, which require trust and detailed hardware knowledge, as in gate-level benchmarking~\cite{Emerson2005}, or additional quantum resources in case we do not have access to or trust in the device to be tested~\cite{GhKa19}.    While our experimental demonstration uses a simplified version~\cite{CEKKZ21} of Mahadev's protocol~\cite{Ma18} we demonstrate the necessary steps for verifying fully untrusted devices. A scaled-up version of our protocol will allow for classical verification, requiring no hardware access or detailed knowledge of the tested device. Its security relies on post--quantum secure trapdoor functions within an interactive proof~\cite{Gold89}. The conceptually straightforward, but technologically challenging scaled-up version of the interactive proofs, considered here, can be used for a variety of additional  tasks such as verifying quantum advantage~\cite{Vidick2021}, generating~\cite{Jacak2021} and certifying quantum randomness~\cite{Bra20}, or composable remote state preparation~\cite{gheorghiu19}.
\end{abstract}

\maketitle

Quantum computers are now widely believed to be at the brink of solving problems that are classically intractable~\cite{Go19,wu2021strong,zhong2020quantum}. Yet, for operating these devices in such a quantum advantage regime, one is confronted with the question of how their output can be verified. The answer depends strongly on the task for which the device is used, the level of control, and its quality. In case the user has direct access to the device, for instance, they can perform gate benchmarks and develop a microscopic error model to gain confidence in the device. For such a scenario, several verification and validation schemes have been proposed~\cite{GhKa19}. However, these techniques are rarely scalable and require detailed hardware knowledge. These requirements can be alleviated somewhat by cross-verifying honest quantum devices~\cite{Elben2020,Greganti2021} to assess their relative performance in a hardware-independent fashion.

The situation becomes significantly more challenging, when the device to be verified cannot be accessed, nor trusted, as might be the case for cloud access quantum computers. Existing techniques that promise to achieve verification without trust require users to have quantum resources~\cite{HM15,GKW15,FK17,ABEM17,TMM19,Barz2012} or two non-communicating quantum processors~\cite{RUV12}. In practice though, a user is typically limited to classical resources. A recent breakthrough showed that even a purely classical user can verify the output of a much more powerful quantum device~\cite{Ma18}. The key idea is to use interactive proofs, where the user exchanges messages with the quantum device to eventually get convinced whether the output is correct, or should be rejected. This approach is secure against dishonest devices under the computational assumption that there exist trapdoor functions that are post--quantum secure~\cite{Re05}. These cryptographic functions have the property that given $f(x)$ it is, even for a quantum computer, not possible to determine the preimage $x$ efficiently. However, having some additional information (trapdoor), the task becomes easy, even for a classical computer. It is widely believed that these functions exist and can be obtained from the Learning With Errors problem~\cite{Re05}. 

Here, we experimentally demonstrate, in a minimal setting, the necessary steps for verifying the output of a quantum computation using only classical resources. We show how to verify existing quantum processors under relaxed security constraints, while the most stringent variant of such a protocol remains too demanding for current quantum hardware. Specifically, we follow the protocol outlined in Ref.~\onlinecite{CEKKZ21} to demonstrate the main ingredients for classical verification tailored to an eight-qubit trapped-ion quantum processor~\cite{Schindler2013}.
Our experiment illustrates that this fully classical verification requires considerably higher fidelity operations of the quantum device than just implementing the underlying quantum computation directly.
Our results below are complementary to the trapped-ion experiment reported recently~\cite{Zhu2021} demonstrating interactive proof protocols with focus on `verifiable quantumness and quantum advantage' (see Appendix~\ref{app:full}).

\begin{figure*}[ht]
\includegraphics[width=\textwidth]{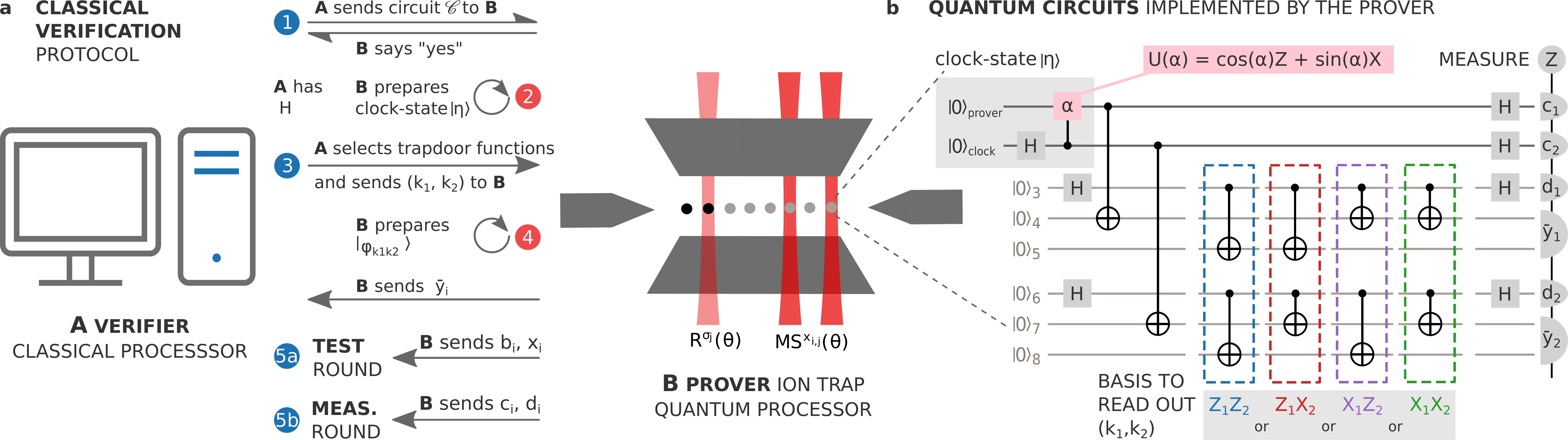}
\caption{\textbf{Classical verification of the output of a quantum computation.}  \textbf{a}, $\sf A$ uses classical resources to verify the answer to a BQP decision problem given by the quantum prover $\sf B$~\cite{Ma18}. To this end, $\sf A$ sends the corresponding circuit $\mathcal{C}$ to $\sf B$ and, depending on $\sf B$'s answer (either ``yes'' or ``no''), constructs a Hamiltonian ($H$) whose ground-state energy encodes the correct output (step 1 in the main text). An honest prover can convince $\sf A$ of the correctness of his answer by efficiently preparing the clock-state $\ket{\eta}$ associated to $\mathcal{C}$ (step 2). Within an interactive proof, $\sf A$ delegates the measurements $(k_1,k_2)$ required to determine the energy of the prepared state to $\sf B$ (step 3). The usage of post--quantum secure trapdoor functions prevents $\sf B$ from learning the measurements he performs (to get $\ket{\phi_{k_1,k_2}}$ according to Eq.\eqref{eq:phi}) and a dishonest prover from convincing $\sf A$ of the wrong output (step 4). To ensure security $\sf A$ requests the measurement outcomes of qubits (4,5,7,8) ($\bar{y}_i$). Once $\sf B$ has committed to an answer, $\sf A$ further requests measurements of qubits (1,3) and (2,6) with probability $1/2$ along $Z$ in a test round (step 5a) to receive $b_i,x_i$, or along $X$ in a measurement round (step 5b) to receive $c_i,d_i$ as indicated in \textbf{b}. This allows $\sf A$ to check whether $\sf B$'s answers are consistent with his initial commitment and to determine the energy. \textbf{b}, Eight-qubit circuit implementing the verification of the outcome corresponding to the single-qubit computation $\mathcal{C}=U(\alpha)$ of Eq.~\eqref{eq:Ualpha}, here using an ion-trap quantum computer. Upon coherent laser-ion interaction single-qubit rotations $\mathsf{R}^{\sigma_j}(\theta)$ and two qubit $\mathsf{MS}^{X_{i,j}}(\theta)$ are realized, see Appendix~\ref{app:setup}. The honest prover prepares the two-qubit clock-state $\ket\eta$ from Eq.~\eqref{eq:eta}, which has low energy with respect to the corresponding two--qubit Hamiltonian. Depending on the verifier's measurement choice, labeled by $(k_1,k_2)$, each qubit of the $\eta$-state is entangled in a particular, for $\sf B$ indecipherable way, to three auxiliary qubits. After measuring some of the qubits in the $Z$-basis, leading to the measurement outcomes $\bar{y}_i$, $\sf A$ chooses randomly to perform either a test or a measurement round. In the former case she checks the correct implementation of $\sf B$ by measuring the remaining qubits in the $Z$-basis, i.e.\ the final Hadamard gates ($H$) are excluded. In the measurement round the remaining qubits are measured in the $X$-basis, which effectively implements a measurement in the $Z_1Z_2$, $Z_1X_2$, $X_1Z_2$ or $X_1X_2$ basis of the qubits in the $\eta$-state. Using the trapdoor information $\sf A$ computes the output of these measurements classically and uses them to determine the energy of the state prepared by $\sf B$. In case the energy is below a certain value $\sf B$ must have been honest.}
\label{Fig:VerificationProtocol}
\end{figure*}

Let us now explain the verification protocol for an arbitrary decision problem within the class of $\sf BQP$ (bounded-error quantum polynomial time), i.e.~a problem which can be solved efficiently by a quantum computer. A description of the corresponding quantum $n$-qubit circuit $\mathcal C$, consisting of $T$ single and two--qubit gates, is sent to the quantum prover Bob ($\sf B$). He can compute the output of the decision problem (``yes'' or ``no'') efficiently. Without loss of generality, we assume that $\sf B$ claims the answer is ``yes'' (the ``no'' case is similar, see Appendix~\ref{app:no}). The classical verifier Alice ($\sf A$) wants to verify this output using only classical means. The correctness of $\sf{B}$'s answer can be checked with the help of an interactive proof. Following Ref.~\onlinecite{KKR05}, they first construct a Hamiltonian $H$ that depends on the prover's answer and the circuit of interest. This Hamiltonian acts on the $n$ system qubits and additionally $\lceil\log(T+1)\rceil$ qubits (the so-called clock register~\cite{KKR05,FeynmanQC}, see Eq.~\eqref{eq:eta}). $H$ can be chosen to consist of only 2-local terms containing the Pauli operators $X$ and $Z$~\cite{KKR05,BL08}. Crucially, the ground-state energy of this Hamiltonian $\la(H)$ encodes the correct output of the initial computation. More precisely, the energy is below a certain value only if $\sf B$'s answer is correct and larger otherwise~\cite{KKR05}. Hence, an honest $\sf B$ can prove that his answer is correct by preparing a state with low energy. To this end, he can prepare efficiently the clock (or history) state $\ket{\eta}$ associated with $\mathcal{C}$~\cite{KKR05}. This state is a superposition, $\sum_{t=0}^T\ket{\Psi_t}\ket{t}$, where $\ket{\Psi_t}$ denotes the state of the system after applying the first $t$ gates of $\mathcal{C}$ to the initial state and the second register denotes the clock register. By construction, this state has low energy, $\bra{\eta}H\ket{\eta}$, in case $\sf B$ is honest as outlined below.
In case $\sf A$ can measure this energy, the problem is solved~\cite{FHM18}. However, $\sf A$ does not possess a quantum device and thus needs to delegate the energy measurement to $\sf B$ without him learning what he is actually measuring. This part of the protocol is achieved using post-quantum-secure trapdoor functions~\cite{Re05} within an interactive proof. $\sf A$ constructs these cryptographic functions and keeps the trapdoor information for herself. This information is what allows her to compute the preimages of the function, which $\sf B$ cannot do efficiently. Furthermore, the functions are of two different types (see below), which will determine whether $\sf B$ performs an $X$ or $Z$ measurement. $\sf B$ cannot efficiently differentiate between the two types of functions and thus cannot learn which measurement he implements by following the measurement protocol depicted in~Fig.\ref{Fig:VerificationProtocol}a. After receiving a description of the function (labeled by $k_i$ in Fig.~\ref{Fig:VerificationProtocol}a), $\sf B$ uses it to entangle each of the qubits in the $\eta$-state with several auxiliary qubits. Some of them are then measured in the computational basis leading to outcomes $\bar{y}_i$ in Fig.~\ref{Fig:VerificationProtocol}. Importantly, the state of the remaining qubits depends on these outcomes. More precisely, the remaining qubits are in a superposition of computational basis states, which are the preimages of $\bar{y}_i$. They are known to $\sf A$. However, $\sf B$ cannot learn them efficiently. At this point the power of interactive proofs comes into play. $\sf B$ is now forced to answer all subsequent questions by $\sf A$ in a way that is consistent with $\bar{y}_i$. 
$\sf A$ could now exploit her superiority to verify quantumness and quantum advantage~\cite{Vidick2021}. In a verification protocol this interactive proof is not only used to ensure that $\sf B$ holds a quantum state, which leads (approximately) to the observed measurement outcomes, but also to enable $\sf A$ to determine its energy in a way hidden to $\sf B$~\cite{Ma18}. Since $\sf B$ can prepare a state with low energy only in case his answer was correct this interactive protocol allows $\sf A$ to verify that answer.

We note that a realization of Mahadev's protocol~\onlinecite{Ma18} requires randomly chosen trapdoor functions with additional properties and with a very large range. 
Hence, many auxiliary qubits are required in the measurement protocol, which is experimentally not feasible on current devices. However, we can demonstrate the key ingredients of classical verification using the simplified version of the protocol outlined in Ref.~\onlinecite{CEKKZ21}. To this end, we use fixed random one-to-one and two-to-one functions~\cite{CEKKZ21}, which map $2$-bit-strings to $2$-bit-strings in each step of the protocol, and we elaborate on further deviations from Mahadev's protocol in Appendix~\ref{app:XZH}.

We implement the protocol experimentally on an eight-qubit trapped-ion quantum processor, see Fig.~\ref{Fig:VerificationProtocol}. On our setup, each $^{40}$Ca$^{+}$ ion resides within a linear string and hosts a single qubit encoded in (meta-)stable electronic states~\cite{Schindler2013}. A universal set of high fidelity quantum gate operations is realized upon coherent laser-ion interaction and comprises arbitrary single-qubit and pairwise two-qubit entangling gates, see Appendix~\ref{app:setup}.

We demonstrate the simplest instance of a verification protocol. Let us denote by  $\mathcal C=U(\al)$ the  single-qubit circuit parametrized by $\alpha$,
\begin{equation}\label{eq:Ualpha}
    U(\al)=\cos\al Z+\sin\al X .
\end{equation}
For convenience, we choose a promise problem with output "yes" if $p_0(\mathcal C) > 3/5$ and "no" if $p_0(\mathcal C) < 1/10$, where $p_0(\mathcal C)=|\bra 0\mathcal C\ket{0}|^2=\cos ^2 \alpha$ (for more details see Appendix~\ref{app:XZH}, in particular, Appendix~\ref{app:no}).
$\sf A$ sends a description of the circuit to $\sf B$, who runs the computation on the quantum computer and obtains an output. Without loss of generality, we assume $\sf B$ claims that the answer is ``yes''. To verify this output the protocol proceeds with the following steps (see Appendix~\ref{app:pro} and Fig.\ref{Fig:VerificationProtocol} for details). 

\textit{Step 1: Determination of the Hamiltonian --} $\sf A$ determines classically the corresponding Hamiltonian~\cite{KKR05} $H$, as given in  Eq.~\eqref{eq:H}. This Hamiltonian acts on two qubits and contains only $X$ and $Z$ operators. 

\textit{Step 2: Preparation of the clock state --} $\sf B$  prepares the clock state corresponding to $U(\al)$ (Fig.\ref{Fig:VerificationProtocol}b), \begin{equation}\label{eq:eta}
  \ket\eta=\frac1{\sqrt2}\left[\ket0\ket0+\left(U(\al)\ket0\right) \ket1\right]\equiv\sum_{b_1,b_2}\alpha_{b_1,b_2}\ket{b_1,b_2}\,.
\end{equation}
Its energy is given by $\bra\eta H\ket\eta=1-p_0(\mathcal C)$ ($= \sin^2 \alpha $). 
Thus, in case the answer of the problem was indeed ``yes'' it holds that $\la(H)<\bra\eta H\ket\eta<2/5$. At the same time we have $\lambda(H)\geq \bra\eta H\ket\eta-2/5$, such that $\la(H)>1/2$ in case the correct answer was ``no" (Appendix~\ref{app:ineq}). 
Consequently, $\sf B$ can only prepare a quantum state with energy below $0.4$ in case his output ``yes'' is indeed the correct output. 

\textit{Step 3: Selection of the trapdoor functions --} In order to delegate the measurements of the operators occurring in the Hamiltonian to $\sf B$, $\sf A$ chooses trapdoor functions $y_k$ labeled by $k=0$ or $k=1$ to perform $Z$ or $X$ basis measurements respectively, see Appendix~\ref{app:pro} for details. We choose the one-to-one function $y_0$ as the identity and the two-to-one function $y_1$ as 
$y_1(z_1,z_2)=(0,0)$ or $(1,0)$ for $z_1=z_2$ or $z_1\neq z_2$ respectively. Here, $z_i \in \{0,1\}$ for $i=1,2$. 
For instance, if
$\sf A$ wants to measure the term $Z_1X_2$, she chooses 
$(k_1,k_2)=(0,1)$. 

\textit{Step 4: Entangling the qubits to be measured with auxiliary qubits --} After receiving $k_1,k_2$, the prover $\sf B$ attaches three auxiliary qubits to each qubit of the
$\eta$-state and implements a unitary operator to generate (depending on the label) one of four 8--qubit entangled states (see Fig.~\ref{Fig:VerificationProtocol}b)
\begin{multline}\label{eq:phi}
  \ket{\phi_{k_1,k_2}}\propto\sum_{b_1,b_2}\al_{b_1,b_2}\ket{b_1}_1\ket{b_2}_2\,\times\\
  \sum_{w_1=0}^1 \ket{w_1}_3\ket{y_{k_1}(b_1,w_1)}_{45}\sum_{w_2=0}^1 \ket{w_2}_6\ket{y_{k_2}(b_2,w_2)}_{78}\,.
\end{multline}

\begin{figure*}[ht]
    \centering
\includegraphics[width=0.9\textwidth]{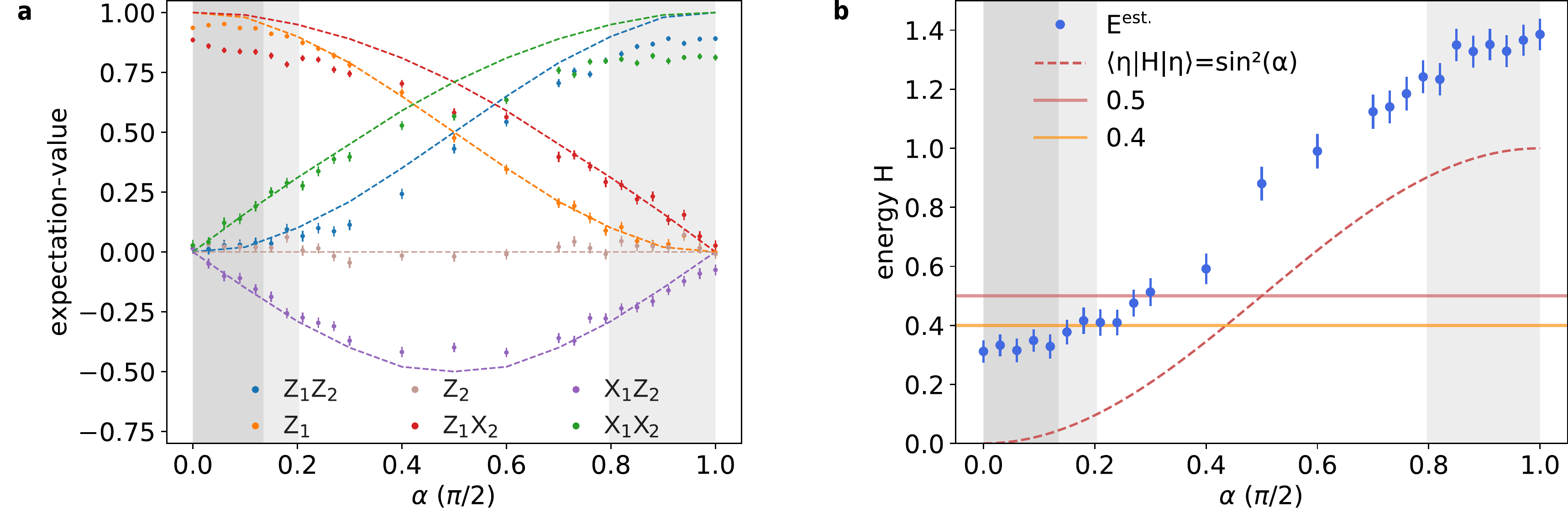}
    \caption{\textbf{Expectation values and energy obtained by $\sf A$ in a measurement round.}
    In a measurement round the qubits of the 
    $\ket\eta$-state are effectively measured in the basis corresponding to $(k_1,k_2)$, $k_i=0$ ($k_i=1)$ meaning that the $i$-th qubit will be measured in the $Z$-basis ($X$-basis) for $i=1,2$ denoting the two qubits of the state $\ket\eta$. To this end, after receiving $\bar y_1$ and $\bar y_2$, $\sf A$ asks $\sf B$ to measure qubits $(1,3)$ and $(2,6)$ in the $X$-basis. Let us denote by $(c_1,d_1)$ and $(c_2,d_2)$ the corresponding measurement outcomes associated with those measurements in qubits $(1,3)$ and $(2,6)$. As explained in Appendix~\ref{app:pro}, given $(c_i,d_i,\bar y_i)$, $\sf A$ can use the trapdoor information to efficiently determine the measurement outcome resulting from projecting the $i$-th qubit of $\ket\eta$ into the $Z$-basis (if $k_i=0$) or the $X$-basis (if $k_i=1$). Note that the measurement round is also used to further get convinced of $\sf B$'s honesty (see Appendix~\ref{app:pro}). \textbf{a}, Accordingly, $\sf A$ obtains expectation values of $XZ$-type operators by choosing $(k_1,k_2)$ appropriately. 
    Dashed lines represent ideal expectation values considering the $\eta$-state (e.g. $\bra\eta Z_1Z_2\ket\eta$ etc.) in the case of a perfect honest prover. \textbf{b}, Total energy $E^{\rm est}$ (with respect to $H$ given by Eq.~\eqref{eq:H}) of the state held by $\sf B$ that is estimated by $\sf A$ when $\sf B$ tries to convince her that the answer was ``yes''. For the problem considered here, the light grey areas are relevant (see Appendix~\ref{app:no}).  Points below $0.4$ (dark grey) allow one to verify that $\sf B$'s claim was indeed correct. In case the answer was ``no'' all states have energy above $0.5$. The upper grey area corresponds to the case where the correct answer was "no", which is also verified here, as explained in Appendix~\ref{app:no}. The red dashed line represents the energy of the $\eta$-state: this is the energy that $\sf A$ would estimate in the ideal case of a perfect, honest prover. All measurements were repeated 2000 times and the error bars represent the standard-deviation from quantum projection noise.}
    \label{Fig:MeasurementRounds}
\end{figure*}

\textit{Step 5: Measurement protocol --} We explain here the measurement protocol for a single qubit, namely qubit 1, in the $\eta$-state, i.e.\ consider the state in Eq.~\eqref{eq:phi} without summing over $b_2,w_2$ and ignoring qubits 2, 6, 7, 8 (the measurement of qubit 2 is equivalent, see  Appendix~\ref{app:pro}). Qubit 1 is entangled in a way that depends on $k_1$ to the three auxiliary qubits, namely qubits 3, 4, 5 (Fig.~\ref{Fig:VerificationProtocol}b). $\sf A$ asks $\sf B$ to measure qubits 4 and 5 in the $Z$--basis and requests the 2-bit outcome, the commitment, (denoted by $\bar{y}_1$ in Fig.~\ref{Fig:VerificationProtocol}). After this
measurement, qubits 1 and 3 are (i) in a product state $\ket{b_1}\ket{x_1}$  where $y_{0}(b_1,x_1)=\bar y_1$ for $k_1=0$; or (ii) in a superposition of the state $\ket0\ket{x_1^{(0)}}$ and $\ket1\ket{x_1^{(1)}}$ where $y_1(0,x_1^{(0)})=y_1(1,x_1^{(1)})=\bar y_1$ for $k_1=1$. Note that $\sf A$ knows which type of function was used and in which state the remaining qubits can be, i.e. she knows the preimage(s) of $\bar y_1$ using the trapdoor information. However, $\sf B$ cannot learn either of them due to the properties of the cryptographic trapdoor functions. This fact implies that $\sf A$ can now ask $\sf B$ for additional measurement outcomes (measuring the remaining qubits) where only she knows the possible outcomes. In order to provide the correct answers, $\sf B$ basically needs to have a quantum state, namely a state of the form from Eq.~\eqref{eq:phi}. Stated differently, the fact that the prover has to commit to an answer ($\bar y_1$) before $\sf A$ asks for additional measurement outcomes, which need to be consistent with the outcome $\bar y_1$ and are unknown to $\sf B$, enforces $\sf B$ to have a quantum state. These properties can be utilized to verify quantumness and quantum advantage, as recently experimentally reported~\cite{Zhu2021}. In the context considered here, the properties explained above also enable $\sf A$ to verify the output of $\sf B$. She proceeds by randomly choosing to perform one of the following two steps. 

\textit{Step 5a: Test round --} $\sf A$ asks $\sf B$ to measure the remaining qubits in the $Z$--basis. The outputs, which are known to $\sf A$ (but not to $\sf B$), need to be consistent with the previous answer. This allows $\sf A$ to test the correct behaviour of $\sf B$ (see Fig.~\ref{Fig:TestinMeasurementRound} for the experimental data and Appendix~\ref{app:full}).

\textit{Step 5b: Measurement round --} $\sf A$ asks $\sf B$ to measure the remaining qubits in the $X$--basis. Given the output, $\sf A$ can determine the measurement outcome corresponding to her choice of $k_1,k_2$ (see Fig.~\ref{Fig:MeasurementRounds} and Appendix~\ref{app:pro}). She uses these outcomes to finally determine the energy. 

Given that $\sf B$ passes the tests of $\sf A$ with sufficiently high probability, a scaled-up version of the protocol (see Appendix~\ref{app:full}) verifies the prover's answer whenever the energy is below $0.4$. Fig.~\ref{Fig:MeasurementRounds}a shows the corresponding experimentally measured expectation values following the above protocol. The data covers the whole  range $0 \leq \alpha \leq \pi/2$ and, up to experimental imperfections, follows the ideal theory prediction (dashed lines). $\sf A$ then makes use of these expectation values to calculate the energy of the state held by $\sf B$ according to the Hamiltonian $H$. Results on the total $\eta$-state energy are depicted in Fig.~\ref{Fig:MeasurementRounds}b, where we successfully certify a ``yes" outcome for $\alpha\leq 0.12\pi/2$. All data-points follow the expected $\braket{\eta\vert H\vert\eta}=\sin^2{\alpha}$ behaviour.
Due to experimental noise, an energy offset appears with respect to the ideal outcome and remains roughly constant over the course of $\alpha$. This is confirmed by a simple depolarizing noise model, outlined in Appendix~\ref{app:noisemodel}, that accurately describes the measured data points.

To emphasize the high control needed over a quantum device to successfully operate the protocol, we quantify the system performance. A simplistic estimate of the fidelity of the experimentally prepared $\ket{\eta}$-state as measured using the six auxiliary qubits is given by considering error-rates on all individual gates in the respective circuits. Considering single- and two-qubit errors inherent to our setup (see Appendix~\ref{app:toolbox}), we expect a fidelity of \SI{0.869(23)}{}. We compare this number to results from Fig.~\ref{Fig:MeasurementRounds}a. For this we use the fact that the $\eta$-state implemented by the circuit in the grey box from Fig.~\ref{Fig:VerificationProtocol}b represents a Bell-state for $\alpha=\pi/2$. Hence, averaging expectation values $Z_1Z_2$ and $X_1X_2$ at $\alpha=\pi/2$ provides an estimate of the $\eta$-state fidelity measured via the six auxiliary qubits, which results in \SI{0.852(8)}{}. This is in good agreement with the above mentioned error-model. Moreover, the latter analysis was similarly performed on the direct estimation of the $\ket{\eta}$-state energy depicted in Fig.~\ref{Fig:EtaStateEnergy} from Appendix~\ref{app:directenergyestimation} and leads to an estimated Bell-state fidelity of \SI{0.945(12)}{}. The difference between expected and observed outcome is due to experimental errors when extracting the $\ket{\eta}$-state energy via the auxiliary qubits as required for the classical verification.
 
To conclude, verification of quantum computation by purely classical means requires both, sophisticated classical tools on the part of the classical user, and large powerful quantum computers on the side of the quantum prover. However, in the continuing effort to relax the constraint to achieve security and in view of wide-ranging and relevant applications~\cite{Zhu2021,Jacak2021,Vidick2021,gheorghiu19}, an inevitable future challenge will be to improve and generalise their sub-protocols both theoretically and technologically, in particular the realisation of interactive proofs using secure post-quantum trapdoor functions. 

\vspace*{5mm}
\noindent
\textbf{Acknowledgment -- } JC and BK are grateful for the support of the Austrian Science Fund (FWF): stand alone project P32273-N27 and the SFB BeyondC F~7107-N38. RS, LP, MM, CE, MR, PS, TM and RB gratefully acknowledge funding by the U.S. ARO Grant No. W911NF-21-1-0007. We also acknowledge funding by the Austrian Science Fund (FWF), through the SFB BeyondC (FWF Project No. F7109), by the Austrian Research Promotion Agency (FFG) contract 872766, by the EU H2020-FETFLAG-2018-03 under Grant Agreement no. 820495, and by the Office of the Director of National Intelligence (ODNI), Intelligence Advanced Research Projects Activity (IARPA), via the U.S. ARO Grant No. W911NF-16-1-0070 and the US Air Force Office of Scientific Research (AFOSR) via IOE Grant No. FA9550-19-1-7044 LASCEM. This project has received funding from the European Union’s Horizon 2020 research and innovation programme under the Marie Sk{\l}odowska-Curie grant agreement No 840450 and No. 801110, and the Austrian Federal Ministry of Education, Science and Research (BMBWF). It reflects only the author's view, the EU Agency is not responsible for any use that may be made of the information it contains. PZ acknowledges support by the US Air Force Office of Scientific Research (AFOSR) via IOE Grant No.~FA9550-19-1-7044 LASCEM, the European Union’s Horizon 2020 research and innovation program under Grant Agreement No. 817482 (PASQuanS), and by the Simons Collaboration on Ultra-Quantum Matter, which is a grant from the Simons Foundation (651440, P.Z.). PZ and RB acknowledge support by the IQI GmbH. \\
\textbf{Appendix -- } is available for this paper.\\
\textbf{Author contributions -- } JC, BK and PZ derived the theory results. RS, MR, LP, MM, CE, PS, and TM performed the experiments. RS analyzed the data. BK and PZ (theory), TM and RB (experiment) supervised the project. All authors contributed to writing the manuscript.\\
\textbf{Competing interests -- } The authors declare no competing interests.\\
\textbf{Materials \& Correspondence -- } Requests for materials and correspondence should be addressed to RS (email: roman.stricker@uibk.ac.at).

\bibliographystyle{apsrev4-1new}
\bibliography{bibliography}

\clearpage
\onecolumngrid
\setcounter{figure}{0}
\setcounter{equation}{0}
\setcounter{table}{0}
\setcounter{section}{0}
\makeatletter 
\renewcommand{\theequation}{A\@arabic\c@equation}
\renewcommand{\thefigure}{A\@arabic\c@figure}
\renewcommand{\thetable}{A\@arabic\c@table}
\renewcommand{\thesection}{A\@Roman\c@section}

\makeatother

\begin{center}
{\bf \large Appendix: \\
Towards experimental classical verification of quantum computation}
\label{SI}
\end{center}
\medskip

\section{Experimental toolbox}
\label{app:setup}
Experiments are performed on an ion-trap quantum computer as illustrated by the middle inset of Fig.~\ref{Fig:VerificationProtocol} from the main text. The setup operates on a linear chain of $^{40}$Ca$^+$ ions confined in ultra high vacuum using a linear Paul trap. Each ion acts as a qubit encoded in the electronic levels $S_{1/2}(m=-1/2) = \ket{0}$ and $D_{5/2}(m=-1/2) = \ket{1}$ denoting the computational subspace~\cite{Schindler2013}. Arbitrary qubit manipulation is realized with coherent laser-ion interaction, upon which the setup is capable of implementing a universal set of quantum gate operations. This universal gate-set comprises of addressed single-qubit rotations with an angle $\theta$ around the x- and the y-axis of the form $\mathsf{R}^{\sigma_j}(\theta) = \exp(-i\theta\sigma_j/2)$ with the Pauli operators $\sigma_j = X_j$ or $Y_j$ acting on the $j$-th qubit, together with two-qubit M\o{}lmer-S\o{}renson entangling gate operations $\mathsf{MS}_{i,j}(\theta) = \exp(-i{\theta} X_i X_j /2)$~\cite{Schindler2013}. Multiple addressed laser beams, coherent among themselves, allow for arbitrary two-qubit connectivity across the entire ion string~\cite{Ringbauer2021}. Initial state preparation in $\ket{0}$ is reached after a series of doppler, polarization-gradient and sideband cooling. Read-out is realized by exciting a dipole transition solely connected to the lower qubit level $\ket{0}$ and collecting its scattered photons, from which the computational basis states $\ket{0}$ and $\ket{1}$ can be identified. Thereby, a qubit's state is revealed by accumulating probabilities from multiple experimental runs. The dipole laser collectively covers the entire ion string, which enables a complete read-out in one measurement round. Additional pump lasers support efficient state preparation as well as cooling and prevent the occupation of unwanted meta-stable states outside the computational subspace $\lbrace\ket{0},\ket{1}\rbrace$.

\section{The protocol step-by-step}
\label{app:pro}
Here we elaborate on steps $1-5$ of the verification protocol discussed in the main text, and as implemented in the experiment.

{\em Step 1: Determination of the Hamiltonian --} 
The Hamiltonian associated to the circuit $\cC=U(\alpha)$ given in Eq.~\eqref{eq:Ualpha} is defined in the Hilbert space of two qubits and reads
$H=H_{\rm out}+6H_{\rm in}+3H_{\rm prop}$ (see Appendix~\ref{app:H}), where
\begin{equation}\label{eq:H}
  \begin{aligned}
    H_{\rm out}&=\tfrac12(1-Z_1-Z_2+Z_1Z_2)\,,\\
    H_{\rm in}&=\tfrac14(1-Z_1+Z_2-Z_1Z_2)\,,\\
    H_{\rm prop}&=\tfrac12(1-\cos\alpha Z_1X_2-\sin\alpha X_1X_2)\,.
  \end{aligned}
\end{equation}

 {\em Step 2: Preparation of the clock state --} As explained in Appendix~\ref{app:ineq}, the ground-energy of $H$ is correlated to the answer of the promise problem and the corresponding clock state (c.f. Eq.~\eqref{eq:eta-gen}) is given in Eq.~\eqref{eq:eta}.

 {\em Step 3: Selection of the trap door function --} Our family of functions contains functions transforming two-bits strings to two-bits strings, and consists of two elements labeled by $k=0$ and $k=1$, respectively. From the main text we recall the definition of the one-to-one function  $y_0(z_1,z_2)=(z_1,z_2)$ (identity), and and two-to-one function $y_1(0,0)=y_1(1,1)=(0,0)$ and $y_1(0,1)=y_1(1,0)=(1,0)$, respectively. $\sf A$ chooses a term $P_1P_2$ in the Hamiltonian that she wants to measure. This determines a pair of labels $(k_1,k_2)\in\{0,1\}^2$ as
\begin{equation}\label{eq:ki}
k_i=\begin{cases}
0,&P_i\in\{Z_i,1_i\}\,,\\
1,&P_i=X_i\,.
\end{cases}
\end{equation}
She sends $(k_1,k_2)$ to $\sf B$. Together with $k$, $\sf A$ generates a trapdoor $t_k$ (see main text) that she keeps private. In the examples considered here, the trapdoor information, $t_k$, together with an output $y_k(z_1,z_2)$ lead to the preimage(s) of $y_k(z_1,z_2)$. 

{\em Step 4: Entangling the qubits to be measured with the auxiliary qubits -- } An honest $\sf B$ prepares the clock state $\ket\eta$ given in Eq.~\eqref{eq:eta} (c.f. Eq.~\eqref{eq:eta-gen}). He attaches  six auxilliary qubits to it, and performs the unitary transformation  $\ket\eta\ket{0^{\otimes 6}}\mapsto\ket{\phi_{k_1,k_2}}$, with $\ket{\phi_{k_1,k_2}}$ given by Eq.~\eqref{eq:phi}. The preparation of those states can be done efficiently by a quantum computer.
 
 {\em Step 5: Measurement protocol --} $\sf B$ is asked to measure some registers of the state he is supposed to hold (the state $\ket{\phi_{k_1,k_2}}$). He is asked to measure the registers $(4,5)$ (obtaining $\bar y_1\in\{0,1\}^2$) and the registers $(7,8)$ (obtaining $\bar y_2\in\{0,1\}^2$)  in the $Z$ basis. He sends $(\bar y_1,\bar y_2)$ to $\sf A$. After these measurements, the state of registers $(1,3)$ and $(2,6)$ depends on the labels $k_1$ and $k_2$, respectively. Let us discuss the cases explicitly:

\begin{itemize}
\item In case $(k_1,k_2)=(0,0)$, the state of the remaining registers will be the product state $\ket{b_1,x_1}_{13}\ket{b_2,x_2}_{26}$, with probability $|\al_{b_1,b_2}|^2$, where $y_0(b_i,x_i)=\bar y_i$ for $i=1,2$.

\item In case $(k_1,k_2)=(0,1)$, the state of the remaining registers will be $\propto\sum_{c=0,1}\al_{b_1,c}\ket{b_1,x_1}_{13}\ket{c,x_{2(c)}}_{26}$, where $y_0(b_1,x_1)=\bar y_1$ and $y_1(0,x_{2(0)})=y_1(1,x_{2(1)})=\bar y_2$.
\end{itemize}
The cases $(k_1,k_2)=(1,0)$ and $(k_1,k_2)=(1,1)$ can be computed in a similar way. After receiving the outcomes $\bar y_1$ and $\bar y_2$, to which $\sf B$ is now committed to, $\sf A$ randomly chooses (with equal probability) to run either a test or a
  measurement round. In both cases, $\sf B$ is asked to measure the remaining registers of the state he is supposed to hold.

{\em Step 5a: Test round --} In a test round, $\sf B$ is
  asked to measure the registers $(1,3)$ and $(2,6)$ in the $Z$ basis and obtains the outcomes $(b_1,x_1)\in\{0,1\}^2$ and $(b_2,x_2)\in\{0,1\}^2$, respectively. He sends the results to $\sf A$. She checks whether $y_{k_i}(b_i,x_i)=\bar y_i$ for $i=1,2$ as would be the case if $\sf B$ was honest. If this is not the case, she rejects. In Fig.~\ref{Fig:TestinMeasurementRound}a we show the probability with which $\sf A$ rejects for each possible label.
  
{\em Step 5b: Measurement round --} In a measurement round, $\sf B$ is
  asked to measure the registers $(1,3)$ and $(2,6)$ , in the $X$ basis and obtains outcomes $(c_1,d_1)\in\{0,1\}^2$ and  $(c_2,d_2)\in\{0,1\}^2$, respectively. He sends the results to $\sf A$. For each qubit $i$, with $i=1,2$, there are two options depending on $k_i$.

\begin{itemize}
\item If $k_i=0$, the $i$th qubit of the clock state was  effectively measured in the $Z$ basis. In this case, $\sf A$ ignores $(c_i,d_i)$ and just computes, with the trapdoor, the preimage of $\bar y_i$ under $y_0(\cdot,\cdot)$. In other words, she finds $(m_i,x_i)$ such that $y_0(m_i,x_i)=\bar y_i$. Finally, she stores $m_i$ as the result of projecting the $i$th qubit in the $Z$ basis.
  
\item If $k_i=1$, the $i$th qubit of the clock state was effectively measured in the $X$ basis. In this case, $\sf A$ stores
  \begin{equation}\label{eq:mi}
  m_i=c_i\oplus d_i\cdot(x_{i(0)}\oplus x_{i(1)})\,,
  \end{equation}
  as the result of projecting the $i$th qubit in the $X$ basis. Here,  $y_{1}(0,x_{i(0)})=y_{1}(1,x_{i(1)})=\bar y_i$. Note that in order to compute $x_{i(0)}$ and $x_{i(1)}$, $\sf A$ needs to use the trapdoor.
\end{itemize}

Summarizing, for each $(k_1,k_2)$, this protocol provides $\sf A$ with a pair of bits $(m_1,m_2)$. By construction, the random variable $(m_1,m_2)$ has the same statistics\footnote{In fact, this is only true if one uses post-quantum secure trapdoor claw-free functions~\cite{Re05} and the prover is accepted in a test round with probability 1. In case this probability is only close to $1$, one can show that the statistics of the measurement outcomes obtained by $\sf A$ are {\em close enough} to those of an actual quantum state. In the general case, this is sufficient to prevent $\sf B$ from cheating (see Appendix~\ref{app:full}).} as the measurement outcomes resulting from projecting the qubits of a quantum state (in case of an honest prover the state $\ket{\eta}$) in the basis associated to $(k_1,k_2)$. Note that the measurement basis is kept secret\footnote{Again, this is true only when considering the family of functions described in~\cite{Ma18}. There it is shown that the labels associated to one-to-one and two-to-one functions are computationally indistinguishable even for a quantum computer (if the problem Learning With Errors is hard for a quantum computer, which is widely believed to be the case~\cite{Re05}).} from $\sf B$. This allows $\sf A$ to estimate the expectation values of $XZ$-type operators corresponding to the state held by $\sf B$ without allowing him to cheat (see Fig.~\ref{Fig:MeasurementRounds}a). Moreover, those expectation values can be used to determine the energy with respect to $H$ (see Fig.~\ref{Fig:MeasurementRounds}b).

\section{The $XZ$-type $\log(n)$-local Hamiltonian and its properties}\label{app:XZH}
Here we give the details of the construction of the Hamiltonian associated to a general decision problem given by a circuit $\cC$ acting on $n$ qubits. One defines the output of the problem to be ``yes'' if $p_0(\mathcal C) > b$ and ``no'' if $p_0(\mathcal C) < a$, where $p_0(\mathcal C)=|\bra{0^n}\mathcal C\ket{0^n}|^2$ and $0\le a<b\le 1$. Considering a uniformly generated family of circuits $\cC_n$ acting on $n$ qubits, one requires $b-a>1/{\sf poly}(n)$. In Appendix~\ref{app:H} we construct this Hamiltonian in case $\sf B$ claims that the answer of the decision problem is ``yes''. In Appendix~\ref{app:ineq} we present the bounds on the ground state energy for the Hamiltonian considered in the main text.  Finally, in Appendix~\ref{app:no}, we explain how this Hamiltonian must be modified in case $\sf B$ claims that the answer of the problem was ``no''.

\subsection{The $XZ$-type $\log(n)$-local Hamiltonian}\label{app:H}
In this section we present the details of the construction of the Hamiltonian $H$ mentioned in the main text associated to an arbitrary circuit $\cC=U_T\cdots U_2U_1$ in case $\sf B$ claims that the answer of the decision problem was ``yes''. This construction is
essentially the same as the one presented in~\cite{KKR05}. However, here our main concern is not the locality of each
term in the Hamiltonian (as shown in the latter reference, it can be made $2$-local) but rather to ensure that the total number of required qubits is kept small. As explained below, to achieve this we will use (a) Gray codes~\cite{PTVF07} and (b) a universal gate set where all the gates are selfadjoint. For a circuit $\cC=U_T\cdots U_2U_1$ acting on $n$ qubits, the Hamiltonian presented in ~\cite{KKR05} $H=H(\cC)$ is acting on $n+\lceil\log(T+1)\rceil$ qubits, with $\lceil . \rceil$ the ceiling function. It  can be expressed as
\begin{equation}\label{eq:H3}
  H=H_{\rm out}+J_{\rm in}\,H_{\rm in}+J_{\rm prop}\, H_{\rm prop}\,,
\end{equation}
where $J_{\rm in}$ and $J_{\rm prop}$ are some suitable polynomials of
$n$. The postive semidefinite operators $H_{\rm in}$, $H_{\rm prop}$ and $H_{\rm out}$ are called
input, propagation and output Hamiltonians, respectively, and will be explicitly presented below. 

The additional $\lceil\log(T+1)\rceil$ qubit register allows us to encode $T+1$ orthogonal quantum states, representing the time steps. Using Gray's code, we write $\ket{t}$ for each of the $T+1$ orthogonal states such
that the representation of two successive values of $t$ differ
only in one bit, i.e., they are given by a Gray code like, for
instance, $(0,1,2,\ldots)=(000,001,011,\ldots)$. The following
expressions, which will be useful in describing the Hamiltonians $H_{\rm out}$, $H_{\rm in}$ and $H_{\rm prop}$, can be written as products of $\log(n)$ operators that are either $X$ or $Z$:
\begin{equation}\label{eqs:Cs}
    C(t)=\,\ket{t}\bra{t}\,,\qquad
    C(t,t-1)=\frac12\,\ket{t}\bra{t-1}+\frac12\,\ket{t-1}\bra{t}\,.
\end{equation}
In fact, the $\lceil\log(T+1)\rceil={\sf O}(\log(n))$-local input and output Hamiltonians are
given by
\begin{equation}\label{eqs:HinHout}
\begin{aligned}
    H_{\rm in}&=\sum_{i=1}^n\frac12({\bf 1}-Z_i)\otimes C(0)\,,\\
    H_{\rm out}&=(T+1)\,\frac12({\bf 1}-Z_1)\otimes C(T)\,,
\end{aligned}
\end{equation}
where the first and second factors in the tensor products act in the
Hilbert space of the $n$ computational qubits and the $\lceil\log(T+1)\rceil$
qubits encoding the clock states, respectively.

The $2$-local Hamiltonian $H_{\rm prop}$ introduced in~\cite{KKR05} can be written as a sum of products of only $X$ and $Z$ operators by using a gadget introduced in Ref.~\cite{BL08}. Importantly, the resulting
Hamiltonian is still $2$-local. However, this comes at the price of
introducing more ancillary qubits. As we show now, using a Gray code and a universal set of self-adjoint gates leads to a Hamiltonian which is no longer $2$--local. However, the number of auxiliary systems is reduced. The result is a
$XZ$-type Hamiltonian that is $\log(n)$-local (instead of just $2$-local). Without loss of generality~\cite{Sh02,BL08}, we can assume that the circuit
$\cC=U_T\cdots U_2U_1$ is written as a sequence of gates $U_i$ that
are either (i) 1-local and of the form $U(\al)=\cos\al Z+\sin\al X$ or
(ii) $\mathsf{CNOT}$ acting on
any pair of qubits. In this case, the ${\sf O}\log(n)$-local propagation Hamiltonian is given by
\begin{equation}\label{eq:Hprop}
\begin{aligned}
  H_{\rm prop}&=\sum_{t=1}^TH_{\rm prop}(t)\,,\\
  H_{\rm prop}(t)&=\frac12{\bf 1}\otimes C(t)+\frac12{\bf 1}\otimes C(t-1)-U_t\otimes C(t,t-1)\,.
\end{aligned}
\end{equation}
One can easily verify that the Hamiltonian $H$ constructed in this
way can be expressed as a sum of products of only $X$ and $Z$
operators, as desired. 

Finally, let us discuss the clock state $\ket\eta$ associated to this Hamiltonian and its relation to its lowest eigenvalue $\la(H)$. The clock state is given by~\cite{KKR05}
\begin{equation}\label{eq:eta-gen}
  \ket\eta=\frac1{\sqrt{T+1}}\sum_{t=0}^TU_t\cdots U_2U_1\ket0\otimes
  \ket t\,,
\end{equation}
and it is easy to see that $p_0(\cC)=1-\bra\eta H\ket\eta$. Hence $\lambda(H)\le\bra\eta H\ket\eta=1-p_0(\cC)$. It is shown in Ref.~\cite{KKR05} that
$\la(H)>\bra\eta H\ket\eta-1/4$ if one chooses appropriate values
$J_{\rm in}={\sf poly}(n)$ and $J_{\rm prop}={\sf poly}(n)$ for the
coefficients in the Hamiltonian. Thus, if
the answer of the problem was ``yes'', $\bra\eta H\ket\eta=1-p_0(\cC)<1-b$ (since $p_0(\mathcal C) > b$ in this case) and thus $\la(H)<\bra\eta H\ket\eta<1-b$ as well.  If the answer of the problem was ``no'', then we have  $\la(H)\ge\bra\eta H\ket\eta-1/4=1-p_0(\cC)-1/4=3/4-a$ (since $p_0(\mathcal C)<a$ in case the answer was``no'').

As mentioned above, in Ref.~\cite{KKR05} a two-local Hamiltonian $H^{\rm (two-loc)}$ is presented which has similar properties to $H$ but is acting on more qubits. There, the coefficients in $H^{\rm (two-loc)}$ can be chosen such that $\la(H^{\rm (two-loc)})<1-b$ in case the answer was ``yes'' and $\la(H^{\rm (two-loc)})>1/2-a$ in case the answer was ``no''~\cite{KKR05}.

In the main text and in what follows, we will choose $J_{\rm in}=6$ and $J_{\rm prop}=3$ (as we are considering here a Hamiltonian acting on a fixed number of qubits). One can see that for our simple example the values $J_{\rm in}=6$ and $J_{\rm prop}=3$ are sufficient to establish~\eqref{ineq:num} while keeping the trace  of the Hamiltonian small enough. This is important because the larger this trace, the larger the impact of errors in the noisy estimation (see the error model given in Appendix~\ref{app:toolbox}). In the next section we explain how, using~\eqref{ineq:num}, the ground-energy of the Hamiltonian~\eqref{eq:H} can be used to encode the original promise problem.

\subsection{Bounds on the ground state energy for the example considered here}\label{app:ineq}
As explained above, the Hamiltonian corresponding to the  circuit $\cC$ encodes the answer to the decision problem ~\cite{KKR05}. Here we give the details for our particular example, where the circuit $\cC=U(\alpha)$ is given in Eq.~\eqref{eq:Ualpha}, the Hamiltonian $H=H_{\rm out}+6H_{\rm in}+3H_{\rm prop}$ is given in Eq.~\eqref{eq:H}, and the clock state is given in Eq.~\eqref{eq:eta}.

For our simple Hamiltonian, one can explicitly check (since it is just a matter
of numerical diagonalization of a $4\times4$ matrix) that
\begin{equation}\label{ineq:num}
  \la(H)>\bra\eta H\ket\eta-2/5\,.
\end{equation}
In our particular case, we take the values $(a,b)=(1/10,3/5)$. So that, if
the answer of the problem was ``yes'', $\bra\eta H\ket\eta=1-p_0(\cC)<1-b=0.4$. In case the answer of the problem was ``no'', we have $\bra\eta H\ket\eta=1-p_0(\cC)>1-a=0.9$ (since $p_0(\mathcal C)<0.1$ in case ``no''). Now, using the latter inequality and Eq.~\eqref{ineq:num}, it
follows that $\la(H)>\bra\eta H\ket\eta-2/5>1-a-2/5=3/5-a=0.5$ in case the answer of the  problem was ``no''. Summarizing, for our particular promise problem, we have 
\begin{equation}\label{eq:sol-la-num}
  \begin{cases}
    \la(H)<\frac25=0.4\,,&{\rm if\,\,\,``yes"}\,,\\[3mm]
    \la(H)>\frac12=0.5\,,&{\rm if\,\,\,``no"}\,.
  \end{cases}
\end{equation}

Recall that we consider here the case where the prover's output is ``yes''. We deal in the next subsection with the case in which he outputs ``no''.

\subsection{Details of the case in which $\sf B$ claims that the answer is "no".}\label{app:no}
As mentioned in the main text, one can assume without loss of generality, that the prover $\sf B$ claims that the answer to the problem associated to the circuit $\cC$ is ``yes''. The reason for that is that in case he claims ``no'' for the circuit $\cC$, this is equivalent to the case where he claims ``yes'' for the modified circuit $\cC'=X\cC$. Here we explain in detail the corresponding Hamiltonian for our particular case.

In case $\sf B$ claims that the answer of the promise problem associated to $\cC$ is ``no'', $\sf A$ constructs a Hamiltonian $H^{\rm (no)}$ in the same way (c.f. Appendix~\ref{app:H}) but now associated to the circuit $X\cC$. Since the circuit always acts on the fixed initial state $\ket0$, one can add an initial $Z$ gate and consider w.l.o.g.\ in the ``no'' case the circuit $X\cC Z$. Since $\cC=U(\al_0)$, one obtains $X\cC Z=U(\pi/2-\al_0)$.

Hence, the situation in which $\sf B$ claims ``no'' for $U(\al_0)$ is equivalent to the situation in which he claims ``yes'' for $U(\pi/2-\al_0)$. For this reason, we study the problem in which the promise is that $\al$ belongs to $I_1\cup I_2$, where $I_1=[0,\arcsin\sqrt{1/10}]$ and $I_2=[\arcsin\sqrt{9/10},\pi/2]$. So that both $\al$ and $\pi/2-\al$ are possible values of our parameter.

A value $\al_0\in I_1\cup I_2$ is given and a description of the circuit $\cC=U(\al_0)$ is sent to $\sf B$. Then the prover $\sf B$ can run this circuit and measure $p_0(\cC)$ to decide whether the answer is either ``yes'' or ``no''. Once $\sf B$ sends his claim (either ``yes'' or ``no'') to $\sf A$, the Hamiltonian $H$ that $\sf A$ constructs is
\[
H=\begin{cases}
H^{\rm (yes)}=H(\cC)=H(U(\al_0)),&{\rm if}\,\,\,{\sf B}\,\,\,{\rm claims}\,\,\,{\rm{``yes"}}\,,\\
H^{\rm (no)}=H(X\cC Z)=H(U(\pi/2-\al_0)),&{\rm if}\,\,\,{\sf B}\,\,\,{\rm claims}\,\,\,{\rm{``no"}}\,,
\end{cases}
\]
where $H(U)$ denotes the Hamiltonian of Appendix~\ref{app:H} associated to circuit $U$. Due to that, the results presented in the main text also show that for any $\al=\pi/2-\al_0$, where $\al_0$ is in the interval for which the experimentally estimated energy is sufficiently low (see Fig.~\ref{Fig:VerificationProtocol}), the same conclusion as for $\al_0$ can be drawn.

\section{Sketch of the fully-secure protocol}
\label{app:full}
The aim of this section is to present more details on the main ingredients, which are required for a completely secure protocol, and to discuss main differences to the simplified version presented in this work. 

In order to have full security, one uses~\cite{Ma18} a family of two-to-one and one-to-one trapdoor functions that are hard to invert even for a quantum computer. Importantly, this family of functions needs to satisfy certain additional technical requirements. First, the two-to-one functions in the family need to have two hardcore bit properties (see Ref.~\cite{Ma18} for details). Roughly speaking, for a two-to-one function $f:\{0,1\}^m\to\{0,1\}^{m'}$ with the hard-core bit property, the following problem is hard: given $x_0\in\{0,1\}^m$ and $f(x_0)=\bar y\in\{0,1\}^{m'}$, find a bit-string $d\in\{0,1\}^m$ such that $d\cdot(x_0\oplus x_1)=0$ (mod $2$), where $f(x_0)=f(x_1)=\bar y$. Second, the one-to-one and the two-to-one functions in the family must {\em look alike}. That is, it must be computationally hard to decide whether a function in the family is two-to-one or not~\cite{Ma18}. As it is unknown how to construct such a family of functions with the required properties, a
family of functions that fulfills those requirements not always but with high probability was used in Ref.~\cite{Ma18}.

\begin{figure}[ht]
\includegraphics[width=0.95\textwidth]{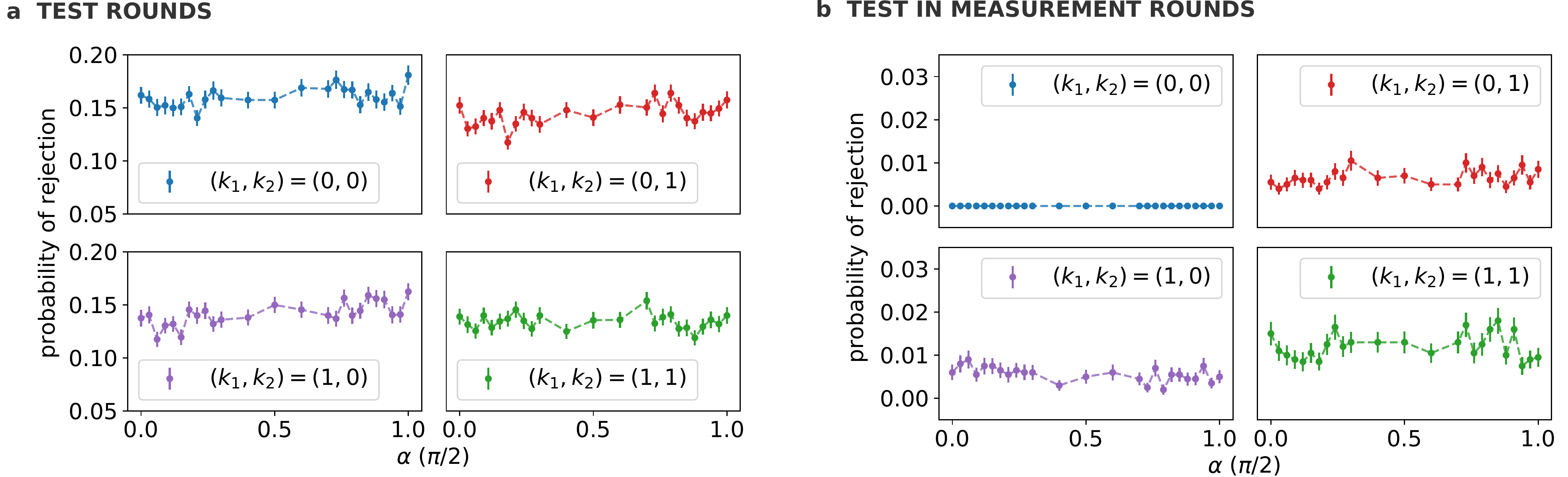}
\caption{\textbf{Probability of rejection in  test and measurement rounds.} \textbf{a}, In a test round $\sf A$ gains confidence that $\sf B$ is holding a quantum state which contains the preimages of $\bar y_i$, as $\sf B$ cannot determine $(b_i,x_i)$ given $\bar y_i$ such that $y_{k_i}(b_i,x_i)=\bar y_i$. To this end, $\sf B$ is asked to measure all the qubits of the state given in Eq.~\eqref{eq:phi} in the $Z$-basis, obtaining $(b_1,x_1,\bar y_1)$ and
$(b_2,x_2,\bar y_2)$ as measurement outcomes, respectively. $\sf A$
checks whether $y_{k_1}(b_1,x_1)=\bar y_1$ and
$y_{k_2}(b_2,x_2)=\bar y_2$. If this is the case, which it would
always be for an honest and ideal prover, $\sf A$ accepts; otherwise she rejects. Here we plot the probability $p_{t,h}$ of $\sf A$ rejecting those outcomes in a single-copy test round for the four possible basis choices $h$, here represented by $(k_1,k_2)$, i.e., the probability that $y_{k_i}(b_i,x_i)\neq\bar y_i$ for some $i=1,2$.
\textbf{b}, In a measurement round, $\sf A$ can perform some additional checks, complementing those of the test round \textbf{a}, by exploiting the fact that the two-to-one functions are not surjective. In our case, this can be easily understood from our choice of the two-to-one
function $y_1:\{0,1\}^2\to\{0,1\}^2$, given by
$y_1(0,0)=y_1(1,1)=(0,0)$ and $y_1(0,1)=y_1(1,0)=(1,0)$. Whenever $k_i=1$, $\sf A$ rejects if $\bar y_i\not\in\{00,10\}$. Observe that in case $k_i=0$, since $y_0(\cdot,\cdot)$ is one-to-one, all possible values $\bar y_i\in\{0,1\}^2$ are possible. Here we plot the single-copy probability of rejection $p_{m,h}$ for the four possible basis choices $h$ represented by $(k_1,k_2)\in\{0,1\}^2$. By construction, this rejection probability is exactly zero for $(k_1,k_2)=(0,0)$. As expected, for $(k_1,k_2)=(1,1)$ this probability is larger than for $(k_1,k_2)=(1,0)$ or $(k_1,k_2)=(0,1)$. Note that this error probability, related to bit-flip errors owed to resonant cross-talk, is in general very small in our experiment (c.f. Appendix~\ref{app:toolbox} and~\ref{app:noisemodel}). Every experimental run was repeated 2000 times. Errors represent 1 standard-deviation from quantum projection noise.}
\label{Fig:TestinMeasurementRound}
\end{figure}

This family of functions is used in Mahadev's measurement  protocol~\cite{Ma18} for both the test and the measurement round. The verifier $\sf A$ decides to run each of these rounds with  equal probability. In both of these rounds, $\sf A$ can reject $\sf B$'s answer. To this end, $\sf A$ uses the trapdoor information and checks whether the preimages of the measurement outcomes $\bar{y}_i$ exist (see Fig.~\ref{Fig:TestinMeasurementRound} for further explanations and the experimental data). 

As explained in the main text, the protocol is used to delegate $X$- and $Z$-basis measurements in a, for $\sf B$ indecipherable way. The statistics of $X$- and $Z$-basis measurements allows to compute the energy of the state with respect to the Hamiltonian $H$, corresponding to an arbitrary decision problem, given in Eq.~\eqref{eq:H}. Its  ground-energy $\la(H)$ encodes the answer to the problem (c.f. Eq.~\eqref{eq:sol-la-num}). In order to determine the energy with respect to $H$, Mahadev uses the protocol presented in~\cite{MNS16}, which we recall here. The Hamiltonian is first written as a convex combination (up to some re--scaling) of projectors, as explained in the following. For a 2--local Hamiltonian  $H=\sum_{l=1}^L c_{l} P(l)$ with $P(l)=P_{i(l)}(l)P_{j(l)}(l)$, where  $P_i(l)\in\{X_i,Z_i,1_i\}$, $i,j$ denote the qubits the operator is acting on and $L={\sf poly}(n)$~\cite{KKR05}, the Hamiltonian $H'=(1+H/c)/2$, where $c=\sum_{l}|c_{l}|$ is defined. The new Hamiltonian $H'$ is a convex combination of projectors of the form $(1+s(l)P(l))/2$ with weights $|c_l|/c$, where $s(l)={\rm sign}(c_l)$. In order to determine the energy, $\sf A$ samples, with probability $|c_{l}|/c$ a term $P(l)=P_{i(l)}(l)P_{j(l)}(l)$, which she then measures. Let $(m_i(l),m_j(l))$ denote the measurement outcomes obtained from projecting qubits $i=i(l)$ and $j=j(l)$ in the eigenbasis of $P_i(l)$ and $P_j(l)$, respectively. We use the notation:

\begin{equation}\label{eq:m}
m(l)=\begin{cases}
(-1)^{m_i(l)+m_j(l)},&{\rm if}\quad P_i(l)\in\{Z_i,X_i\}\,,\quad P_j(l)\in\{Z_j,X_j\}\,,\\
(-1)^{m_i(l)},&{\rm if}\quad P_i(l)\in\{Z_i,X_i\}\,,\quad P_j(l)=1_j\,,\\
(-1)^{m_j(l)},&{\rm if}\quad P_j(l)\in\{Z_j,X_j\}\,,\quad P_i(l)=1_i\,,\\
1,&{\rm if}\quad P_i(l)=1_i\,,\quad P_j(l)=1_j\,.
\end{cases}
\end{equation}
Now, $\sf A$ uses the following rule to determine a final bit $r(l) \in\{0,1\}$:
\begin{equation}\label{eq:r}
r(l)=\begin{cases}
1,&{\rm if}\quad m(l)=s(l)\,,\\
0,&{\rm if}\quad m(l)=- s(l)\,.
\end{cases}
\end{equation}
Observe that if $\sf B$ could send a state to $\sf A$, and $\sf A$ could perform those measurements by herself, the expected value $\langle r\rangle$ of the random variable $r$ coincides with the energy, with respect to $H'$, of the state sent by $\sf B$~\cite{MNS16}. Hence, after repeating the measurement protocol $N={\sf poly}(n)$ times obtaining $\{r_1,\ldots,r_N\}$, $\sf A$ can compute an estimate $r^{\rm est}$ of the expectation value  $\langle r\rangle$ of the random variable $r$. She then ``accepts'' if $r^{\rm est}\leq T_0$, with
\begin{equation}\label{eq:T0}
T_0=(1+f(H)/c)/2\,,
\end{equation}
where $f(H)$ denotes  the upper bound on $\la(H)$ given in Appendix~\ref{app:H} in case the answer was ``yes''. For the example considered in the main text we have 
$T_0=(1+0.4/c)/2$ using Eq.~\eqref{eq:sol-la-num}. In case the answer to the problem was ``no'', one has $r^{\rm est}\ge T_1$, with
\begin{equation}\label{eq:T1}
T_1=(1+g(H)/c)/2\,.
\end{equation}
where $g(H)$ denotes the lower bound on $\la(H)$ given in Appendix~\ref{app:H} in case the answer was ``no''. For the example considered in the main text we have 
$T_1=(1+0.5/c)/2$ using Eq.~\eqref{eq:sol-la-num}.

As the two bounds, $T_0,T_1$ differ by $1/{\sf poly}(n)$, one can run an extended protocol, using polynomially many copies $N$ of the state to differentiate between the two cases (``yes'' and ``no'') with an exponentially small error (see~\cite{FHM18} and Protocol 8.3. in~\cite{Ma18}).  

Note that for convenience in our experiment we do not estimate $\langle H'\rangle$ by sampling from the probability distribution $|c_l|/c$, but determine $\langle H\rangle$ instead. This amounts to distinguishing energies below $0.4$ and above $0.5$ (c.f. Fig.~\ref{Fig:MeasurementRounds}). In the protocol discussed here, one needs to distinguish quantities below $0.4/c$ and above $0.5/c$. However, using several repetitions of the protocol would allow us to distinguish the two cases. In particular, in our example, for $\al=0.12\pi/2$ one has $c=12.4631$.

A classical verifier $\sf A$ uses Mahadev's measurement protocol as follows to delegate the previous measurements to $\sf B$ for the extended protocol.

\begin{itemize}
\item 
First, she randomly chooses  $N={\sf poly}(n)$ operators $P_1,\ldots, P_N$ occurring in the Hamiltonian $H'$  independently with probability $\{|c_l|/c\}_l$. For each choice, $l$, she defines a vector $h^l\in \{0,1\}^n$ by setting 
\[
h^l_i=\begin{cases}
0,&{\rm if}\quad P_i\in\{Z_i,1_i\}\,,\\
1,&{\rm if}\quad P_i=X_i\,.
\end{cases}
\]

Note that $h^l_i=0$ for any qubit $i$ on which $P_l$ acts trivially.  We call the vector $h=(h^1,h^2, \ldots, h^N)$ now the basis choice. 

\item $\sf A$ and $\sf B$ run the measurement protocol explained in the main text for the basis choice $h$. Let $p_{t,h}$ ($p_{m,h}$) denote the probability that at least one of the tests in the test (measurement) round failed. 

\item In the measurement round the verifier computes the product of the measurement results for each term $P(l)$ and sets $r(l)=1$ only if for more than half on the times the product of the measurement results coincides with $s(l)$.

\end{itemize}

Given $r(l)$, $\sf A$ computes the expectation value of $r$ and thereby the expectation value of $H'$. As shown in \cite{Ma18} the probability with which $\sf A$ accepts the answer of $\sf B$ is given by 

\begin{equation}\label{eq:pa}
P_{\rm accept}=\frac12\sum_h v_h(1-p_{t,h})+\frac12\sum_h v_h(1-p_{m,h}){\sf Prob}_h(r^{\rm est}<T_0)\,.
\end{equation}
Here, $v_h$ denotes the probability with which $\sf A$ samples the basis choice $h$ (depending on the Hamiltonian). Moreover, ${\sf Prob}_h(r^{\rm est}<T_0)$ denotes the probability with which $r^{\rm est}<T_0$ in case the basis choice $h$ was used for the measurements. 

Note that, in the absence of noise, an honest prover $\sf B$, who would simply prepare $N$ copies of the $\ket{\eta}$--state, would be accepted with probability exponentially close to $1$~\cite{Ma18}. Importantly, in Ref.~\cite{Ma18}, Mahadev showed that $\sf A$ can differentiate between such a honest prover and a dishonest prover. The reason for that is that the probability for accepting a dishonest prover is upper bounded by $3/4$ 
even for non-vanishing $p_{t,h}$ and $p_{m,h}$ (soundness). Repeating the protocol ${\sf poly}(n)$ many times the verifier can distinguish between the cases where the outcome of the problem was ``yes'' or ``no''. 

Let us also finally comment on other applications of interactive proofs
using post-quantum secure cryptographic functions. They can be used to verify quantumness and quantum advantage~\cite{Bra20,Vidick2021}, which has been recently demonstrated experimentally~\cite{Zhu2021}. ${\sf A}$ can use the above ideas to get convinced that ${\sf B}$ possesses a quantum state and is using it to solve certain computational task efficiently, which would have been impossible for a classical machine. In particular, $\sf A$ can send $\sf B$ the labels $k$ of two-to-one trapdoor claw-free functions $F_k:\{0,1\}^m\mapsto\{0,1\}^{m'}$ and ask him to prepare the state
\[
\frac1{\sqrt{2^m}}\sum_{x=0}^{2^m-1}\ket x\ket {F_k(x)}\,.
\]
Then, $\sf A$ asks $\sf B$ for a commitment string $\bar y\in\{0,1\}^{m'}$ in the range of $F_k$. This can be easily provided by a quantum  $\sf B$. He just needs to measure the last $m'$ registers. Then, $\sf B$ would hold the superposition
\[
\frac1{\sqrt2}\ket{x_0}+\frac1{\sqrt2}\ket{x_1}\,,
\]
where $F_k(x_0)=F_k(x_1)=\bar y$. Recall that $\sf B$ does not know the preimages, $x_0,x_1$. However, $\sf A$ can easily compute them knowing the trapdoor information. Now, $\sf A$ asks $\sf B$ to measure the remaining $m$ qubits either in the $Z$- or $X$-basis and to send the result to $\sf A$. We denote by $p_A$ the probability that $\sf B$ sends a preimage of $\bar y$ in the first case. An honest $\sf B$ would just measure the first $m$ registers in the $Z$-basis and obtain a bit-string $x_b$ such that $F_k(x_b)=\bar y$, where $b=0$ or $b=1$ with probability $1/2$. Let $p_B$ denote the  probability that $\sf B$ sends $d\in\{0,1\}^m$ such that $d\cdot(x_0\oplus x_1)=0$ (mod $2$) with $F_k(x_0)=F_k(x_1)=\bar y$, in the second case. It can be easily seen that, in this case, an honest $\sf B$ would just measure the first $m$ registers in the $X$-basis and obtain a string $d$ that would be accepted. Note that in case of $X$-basis measurements, the hard-core bit property discussed above comes into play. Recall that the function $F_k$ has  the hard-core bit property if, given $x_0\in\{0,1\}^m$ and $F_k(x_0)=\bar y\in\{0,1\}^{m'}$, it is hard to find a bit-string $d\in\{0,1\}^m$ such that $d\cdot(x_0\oplus x_1)=0$ (mod $2$), where $F_k(x_0)=F_k(x_1)=\bar y$. Note that, in contrast to a classical device, an honest quantum $\sf B$ would be able to obtain such $d$ without knowing $x_0$ or $x_1$. Using these properties, one can derive interactive proof protocols based on the assumption that the problem Learning With Errors is hard~\cite{Re05} such that $p_A+2p_B\le 2$ for the best classical strategy (for $m,m'$ sufficiently large). This means that such a protocol can be used by $\sf A$ to verify the ``quantumness'' of $\sf B$ and even quantum advantage in case $m,m'$ are large enough~\cite{Vidick2021,Bra20,Zhu2021}.

\section{Ion-trap implementation and error-rates}
\label{app:toolbox}
\begin{figure}[ht]
\includegraphics[width=0.7\textwidth]{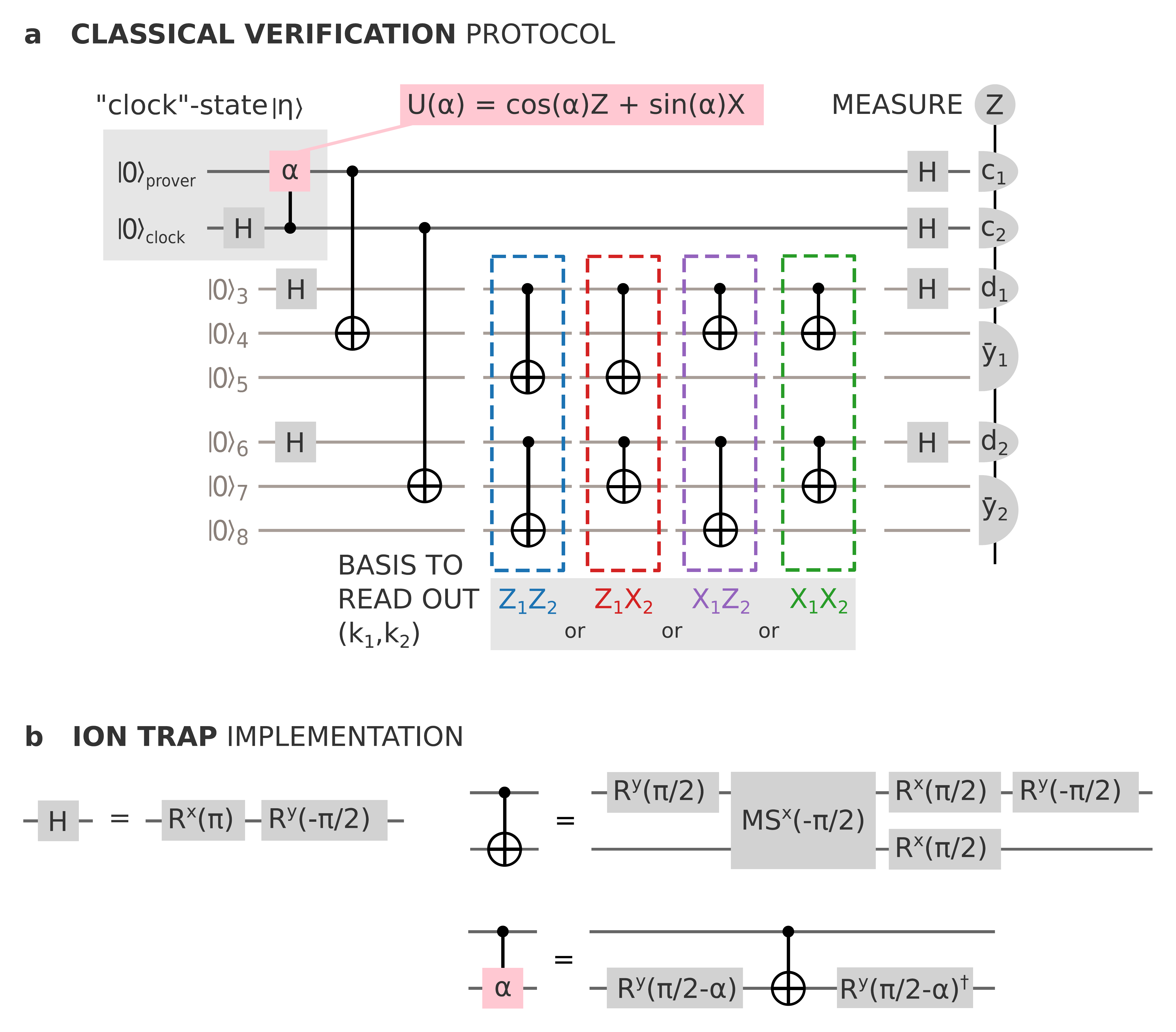}
\caption{\textbf{Ion trap implementation of the classical verification protocol.} \textbf{a}, Eight-qubit circuit implementing our decision problem linked to the outcome of the gate $U(\alpha)=\cos\alpha Z+\sin\alpha X$, according to Fig.~\ref{Fig:VerificationProtocol} from the main text. The circuit consists of single qubit gates, i.e. Hadamards $\mathsf{H}$, as well as $\mathsf{CNOT}$-gates to create pairwise entanglement. The final read-out in Z-basis is performed via collective fluorescence detection, see text. \textbf{b}, Efficient ion-trap implementation of the building blocks from \textbf{a}. We emphasize that one $\mathsf{CNOT}$ requires one two-qubit $\mathsf{MS}^{X_{i,j}}(-\pi/2)$ alongside four single-qubit gates. We implement $\mathsf{CU(\alpha)}$ using two additional single-qubit gates on the prover-qubit to continuously change between $\mathsf{CPHASE}$ and $\mathsf{CNOT}$ for $\alpha=0$ and $\pi/2$ respectively. Compiling the final circuits results in 19 single-qubit gates alongside five $\mathsf{MS}^{X_{i,j}}(-\pi/2)$.}
\label{Fig:ClassicalVerificationCircuit}
\end{figure}

The particular circuit for the classical verification protocol discussed in the main text is again depicted in Fig.~\ref{Fig:ClassicalVerificationCircuit}a with a focus on the ion-trap implementation. The circuit demands for local gates, more specifically Hadamards $\mathsf{H}$, as well as two-qubit $\mathsf{CNOT}$-gates - the latter creating pairwise entanglement. Fig.~\ref{Fig:ClassicalVerificationCircuit}b follows up on the sub-circuits corresponding to those building blocks suitable and optimized for the ion-trap gate set. Each $\mathsf{CNOT}$ gate demands for a full-entangling, two-qubit $\mathsf{MS}^{X_{i,j}}(-\pi/2)$ alongside four single-qubit gates, i.e. single-qubit rotations of type $\theta=\pi/2$ around X, Y or Z. $\mathsf{CU(\alpha)}$ from the grey box is realized upon two single-qubit gates acting on the prover-qubit to continuously change basis between $\mathsf{CPHASE}$ and $\mathsf{CNOT}$ for $\alpha=0$ and $\pi/2$ respectively. The total number of single-qubit gates is further reduced by compiling the overall circuit. Thus, the final implementation of each circuit $Z_1Z_2$, $Z_1X_2$, $X_1Z_2$ and $X_1X_2$ requires five $\mathsf{MS}^{X_{i,j}}(-\pi/2)$ alongside 19 single-qubit gates ($\theta=\pi/2$ around X, Y or Z).

Note that all results from the main text, covered by Fig.~\ref{Fig:MeasurementRounds} have been accumulated from 2000 experimental runs in each data point to faithfully estimate the protocol's outcome. The respective number of experimental runs in complementary experiments covered by this Appendix have further been stated in the individual figure captions. Generally, for the estimation of quantum projection noise, as stated by error-bars in figures and errors in numbers, the probabilities of measured outcomes were resampled using a multinomial distribution considering the number of experimental runs. If not stated differently the underlying errors were then extracted from the resampled data-set and correspond to 1 standard deviation.

In the following we discuss error-rates inherent to our system. Our average single-qubit fidelity ($\theta=\pi/2$ around X, Y or Z) estimated via randomized benchmarking reads \SI{0.9994(3)}{}~\cite{Ringbauer2021}. To further improve single-qubit gates on circuit level, we construct each gate out of three gates using various axes following the aim of reducing cross-talk to neighbouring qubits. This results in a slightly lower average fidelity on the composite gate of \SI{0.998(1)}{}. However, on the eight-qubit circuits this approach is beneficial, as otherwise cross-talk errors proliferate generally lowering the implementation's quality.

\begin{figure}[ht]
\includegraphics[width=0.45\textwidth]{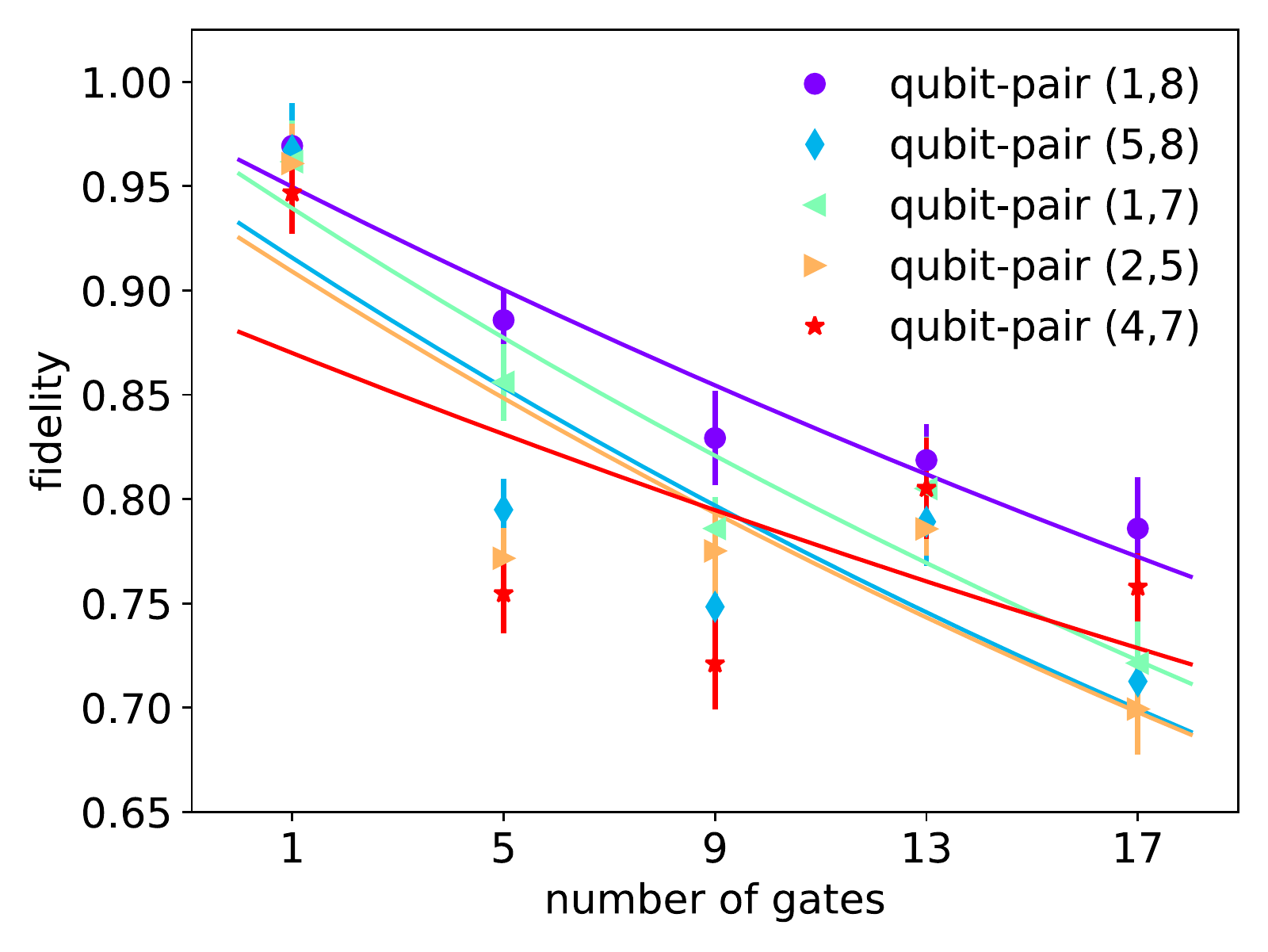}
\caption{\textbf{Error-rates on pairwise $\mathsf{MS}$-gates from the $X_1X_2$ measurements}. For each qubit-pair $(i,j)$, a GHZ-state was prepared from a series of $4n+1$ with $n\in\lbrace0,1,2,3,4\rbrace$ full-entangling $\mathsf{MS}^{X_{i,j}}(-\pi/2)$. Fidelities are then inferred from population, i.e. the probability of measuring states $p_{00}$ and $p_{11}$ as well as from coherence represented by the contrast in parity oscillations, see text for details. Each data-point represents 100 experimental runs. Error-bars correspond to 1 standard deviation due to quantum projection noise. Overall fidelities fitted from the exponential decay are further summarized in Tab.~\ref{SITab:MSGateDecay}.}
\label{Fig:MSGateDecay}
\end{figure}
 
The performance of the two-qubit $\mathsf{MS}$-gates may slightly differ upon the chosen qubit-pair along the ion-string. To give an estimate, we characterize the particular five pairs occurring in the $X_1X_2$-measurement. Note that the qubit order differs from the upper circuits, as in the actual ion-trap implementation we optimized for inter ion spacing to keep cross-talk as low as possible. Given the similar ion-pairs for the $Z_1Z_2$, $Z_1X_2$, $X_1Z_2$ measurements this is a representative set for all circuits. Results are depicted in Fig.~\ref{Fig:MSGateDecay}. For each pair a GHZ-state is initialized through a series of $(4n+1)$ with $n\in\lbrace0,1,2,3,4\rbrace$ full-entangling $\mathsf{MS}^{X_{i,j}}(-\pi/2)$. Here we use that the GHZ-state's density matrix ideally consists of only four elements, i.e. two diagonal elements $\ket{00}$ and $\ket{11}$ referred to as population as well as two off-diagonal elements corresponding to relative coherence. The population can be directly inferred from fluorescence detection of the measured population in $p_{00}$ and $p_{11}$, whereas the coherence terms are extracted from the contrast in parity oscillations. Averaging population and parity leads to the GHZ-state fidelity. Final numbers, extracted from the $\mathsf{MS}$-gate series' decay, scatter between \SI{0.976(9)}{} and \SI{0.984(12)}{}. Hence the overall performance of our implementation will be limited by $\mathsf{MS}$-gates rather than by single-qubit errors. Moreover, all results on population, coherence and fidelity are summarized in Tab.~\ref{SITab:MSGateDecay}. We emphasize that similar results on population and parity decay led us to conclude that the $\mathsf{MS}$-gate performance is generally dominated by depolarizing noise, which we will make use of in an error-modelling to characterize system limitations - thoroughly discussed at the bottom in Appendix~\ref{app:noisemodel}.

\begin{table}[ht]
\centering
\caption{\textbf{Summary of error-rates on $\mathsf{MS}$-gate pairs according to Fig.~\ref{Fig:MSGateDecay}}. Results on population, coherence and fidelity are extracted from the exponential decay of a series of $\mathsf{MS}$-gates as depicted in the figure. All errors refer to  1 standard deviation from the exponential fit uncertainty.}
\begin{tabular}{c r r r r r}
    \toprule
    \midrule
    \multicolumn{1}{c}{$\mathsf{MS}$-gate} & \multirow{1}[2]{*}{} & \multicolumn{1}{c}{population} & \multicolumn{1}{c}{coherence} &  \multirow{1}[2]{*}{} &\multicolumn{1}{c}{fidelity}\\ 
    \cmidrule(rl){1-2}\cmidrule(rl){3-4}\cmidrule(rl){5-6}
    qubit-pair & & $\mathcal{F}_\text{pop.}$ & $\mathcal{F}_\text{coh.}$ & & $\mathcal{F}_\text{tot.}$ \\
    \cmidrule(rl){1-2}\cmidrule(rl){3-4}\cmidrule(rl){5-6}
    \multicolumn{1}{r}{(1,8)}  &  & 0.979(3) & 0.983(3) & & 0.982(3) \\
    \multicolumn{1}{r}{(5,8)}  &  & 0.981(14) & 0.973(6) & & 0.976(8) \\
    \multicolumn{1}{r}{(1,7)}  &  & 0.974(5) & 0.978(5) & & 0.977(5) \\
    \multicolumn{1}{r}{(2,5)}  &  & 0.974(14) & 0.977(6) & & 0.976(9) \\
    \multicolumn{1}{r}{(5,8)}  &  & 0.983(16) & 0.984(10) & & 0.984(12) \\
    \midrule
    \bottomrule
\end{tabular}
\label{SITab:MSGateDecay}
\end{table}

We work these error-rates into a simplistic estimate on the expected fidelity of the $\eta$-state as measured using the six auxiliary qubits. Therefore, we accumulate error-rates on all gates from the circuits. The 19 single-qubit gates reduce the fidelity to \SI{0.966(18)}{}. Additionally taking $\mathsf{MS}$-gate rates from Tab.~\ref{SITab:MSGateDecay} into account, we expect a final fidelity of
\[
\mathcal{F} = 0.966(18)\cdot0.982(3)\cdot0.976(8)\cdot0.977(5)\cdot0.976(9)\cdot0.984(12) = 0.869(23)
\]
for the $\eta$-state measurement. We compare this to the results from the experimental implementation depicted in Fig.~\ref{Fig:MeasurementRounds}a by averaging $Z_1Z_2$ and $X_1X_2$ at $\alpha=\pi/2$ representing an estimate of the $\eta$-state's Bell-state fidelity. The result reads:
\[
\mathcal{F} \sim \frac{Z_1Z_2(\alpha=\pi/2)+X_1X_2(\alpha=\pi/2)}{2} = \frac{0.891(10)+0.812(13)}{2} = 0.852(8),
\]
and is again in good agreement with the above simplistic error-modelling.

\section{A different decision problem}
\label{app:alphato1}
Here we give the details of our experimental results for a different, but related, decision problem. Specifically we consider the problem where the answer is ``yes'' for $\alpha$ close to $\pi/2$ in the same circuit $\cC=U(\al)$ as in the main text. Crucially, from a computer science perspective this case is completely equivalent to the case discussed in the main text. In practice, however, different decision problems, even when they are associated to the same circuit, correspond to different implementations at the hardware level and might thus exhibit different noise sensitivity. For this reason, it might be interesting to verify multiple instances on a given quantum device. To confirm this behaviour, we now consider the circuit 
$\cC=U(\al)=\cos\al Z+\sin\al X$, but defining the answer of the problem to be
\begin{equation}\label{eq:sol-p1-new}
  \begin{cases}
    {\rm{``yes"}}\,,&{\rm if\,}p_1(\mathcal C) > b\,,\\
    {\rm{``no"}}\,,&{\rm if\,}p_1(\mathcal C) < a\,,
  \end{cases}
\end{equation}
under the promise that one of the two cases occurs, where $p_1(\cC)=|\bra 1\cC\ket 0|^2$. The corresponding Hamiltonian $H=H_{\rm out}+J_{\rm in}H_{\rm in}+J_{\rm prop}H_{\rm prop}$ (see Appendix~\ref{app:H}) must thus be changed such that $H_{\rm out}$ penalizes states with the output-qubit in state $\ket 0$, such that
\[
H_{\rm out}=(T+1)\frac12({\bf 1}+Z_1)\otimes C(T)\,.
\]
Assuming now, without loss of generality (see Appendix~\ref{app:no}), that $\sf B$ claims that the answer to the decision problem was ``yes'', the Hamiltonian $H=H_{\rm out}+6 H_{\rm in}+3 H_{\rm prop}$ now reads
\begin{equation}\label{eq:H-new}
  \begin{aligned}
    H_{\rm out}&=\tfrac12(1+Z_1-Z_2-Z_1Z_2)\,,\\
    H_{\rm in}&=\tfrac14(1-Z_1+Z_2-Z_1Z_2)\,,\\
    H_{\rm prop}&=\tfrac12(1-\cos\al Z_1X_2-\sin\al X_1X_2)\,,
  \end{aligned}
\end{equation}
to be compared with the Hamiltonian of Eq.~\eqref{eq:H} in the main text. Analogously to the case in the main text, one can show that in this case $\bra\eta H\ket\eta=1-p_1(\cC)$ and $\la(H)>\bra\eta H\ket\eta-2/5=3/5-p_1(\cC)$ hold (see Appendix~\ref{app:ineq}). This implies that Eq.~\eqref{eq:sol-la-num} holds for the Hamiltonian~\eqref{eq:H-new} and the decision problem~\eqref{eq:sol-p1-new}. 

\begin{figure*}[ht]
    \centering
    \includegraphics[width=0.9\textwidth]{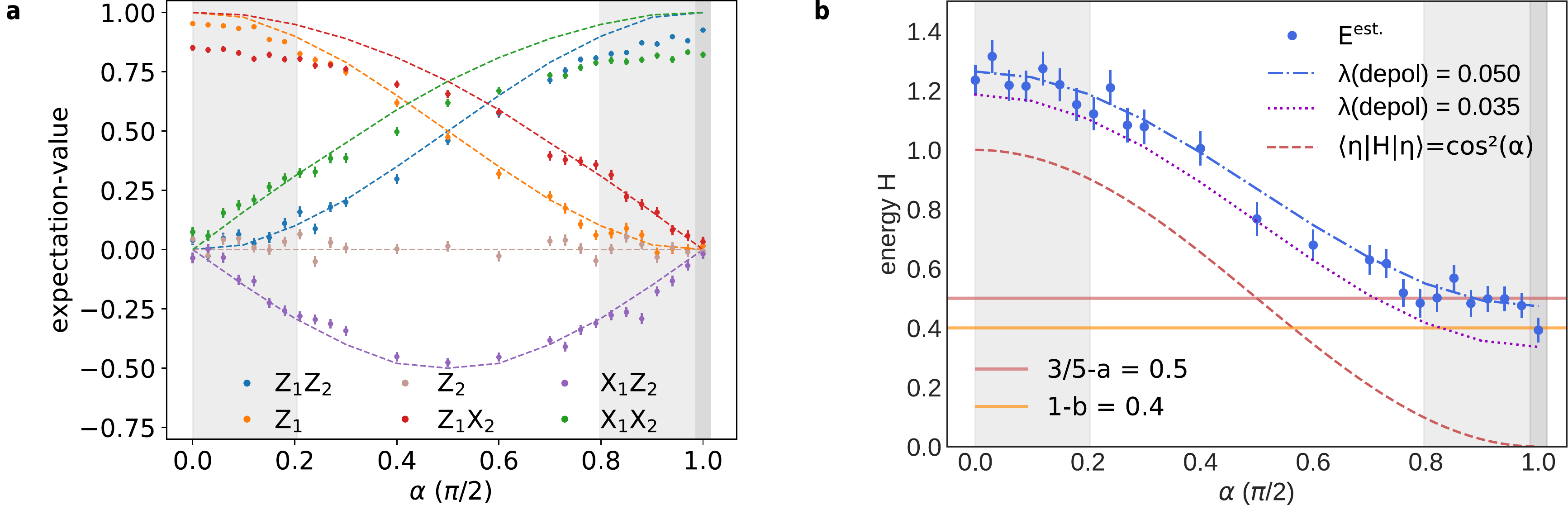}
    \caption{\textbf{Expectation values and energy obtained by $\sf A$ in a measurement round for the new decision problem from Eq.~\eqref{eq:sol-p1-new}.}
    \textbf{a}, Experimentally received operator values similar to Fig.~\ref{Fig:MeasurementRounds} from the main text. The results were obtained after repeating the measurement round of the protocol 2000 times for each value $(k_1,k_2)\in\{0,1\}^2$. Note that the definition of our decision problem $(a,b)=(1/10,1/5)$ implies only $\alpha\in[1.25,1.57]$ (light grey) to be relevant. Dashed lines show ideal outcomes. \textbf{b}, Total energy of the state hold by $\sf B$ estimated by $\sf A$ for the Hamiltonian Eq.~\eqref{eq:H-new}. We again find depolarizing noise (dashed-dotted-line) at a rate of $\lambda=0.05$ well describing our noisy data. Further, a 30\% reduction in system noise is sufficient to successfully operate the protocol, as indicated by the dotted-line at $\lambda=0.035$, see Appendix~\ref{app:noisemodel} for details. Errors represent 1 standard-deviation from quantum projection noise.}
    \label{Fig:MeasurementRoundsAlpha1}
\end{figure*}

Figure~\ref{Fig:MeasurementRoundsAlpha1} shows the experimental results in this case, again choosing $(a,b)=(1/10,3/5)$ so that the thresholds of Eq.~\eqref{eq:sol-la-num} remain $1-b=0.4$ and $3/5-a=0.5$. Curiously, despite being formally equivalent to the case in the main text, verification turns out to be slightly more challenging for this problem. This goes to show, that the protocol indeed verifies the output of a device, not the device itself. Hence, just because the protocol successfully verifies one instance, does not mean that all instances can be verified.

A closer inspection of the underlying circuits show that the case $\alpha \to \pi/2$ generates more entanglement in the system compared to $\alpha \to 0$. Experimentally, this amplifies the noise in an unfavourable way to prevent verification for most values of $\alpha$ in this case. One exception is the instance with $\alpha=\pi/2$ which features an energy below the verification threshold for two reasons. First, the term given in the above Hamiltonian proportional to $\cos{\alpha}Z_1X_2$ exactly vanishes at $\alpha=\pi/2$. Second, the basis change operation rotating $\mathsf{CNOT}$ into $\mathsf{CPHASE}$ becomes trivial for $\alpha=\pi/2$. We note that the same depolarizing noise model used in the main text (see Appendix~\ref{app:noisemodel}) also accurately describes the data presented in Fig.~\ref{Fig:MeasurementRoundsAlpha1} here. Numerical simulations then suggest that about a 30\% reduced depolarizing rate of $\lambda=0.035$ would be required to verify this case. This demonstrates that there can be large differences in the verifiability of different instances of the same problem on the same hardware.

\section{Direct estimation of $\eta$-state energy}
\label{app:directenergyestimation}
\begin{figure*}[ht]
    \centering
    \includegraphics[width=\textwidth]{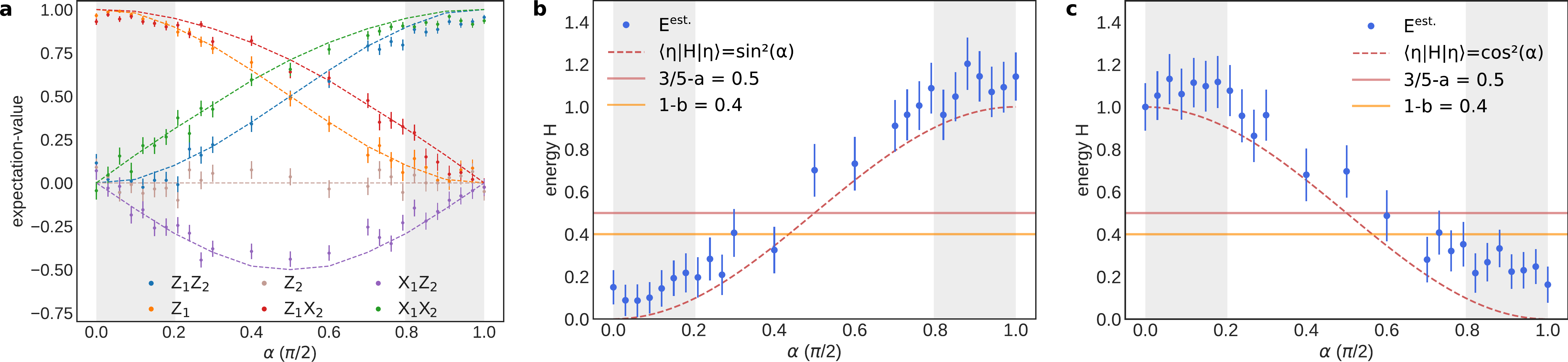}
    \caption{\textbf{Direct estimation of $\eta$-state total energy.} Preparation of solely the $\eta$-state according to the circuit depicted in the grey box from Fig.~\ref{Fig:ClassicalVerificationCircuit}. \textbf{a}, Directly on $\eta$-state qubits measured expectation values for each value $(k_1,k_2)\in\{0,1\}^2$ covering $0 \leq \alpha \leq \pi/2$. \textbf{b}, Total energy of the $\eta$-state as in the decision problem of the main text relating $\alpha\rightarrow0$ as the ``yes'' outcome - calculated from results in \textbf{a}. \textbf{c}, Total energy of the $\eta$-state as in the decision problem from Appendix~\ref{app:alphato1} relating $\alpha\rightarrow\pi/2$ as the ``yes'' outcome - calculated from results in \textbf{a}. Each experimental run was repeated 400 times. Errors represent 1 standard-deviation from quantum projection noise. Dashed lines follow the ideal outcome.}
    \label{Fig:EtaStateEnergy}
\end{figure*}

For previous attempts, covered in Fig.~\ref{Fig:MeasurementRounds}b and Fig.~\ref{Fig:MeasurementRoundsAlpha1}b, the $\eta$-state's energy was estimated using the six auxiliary qubits necessary for the trapdoor function implementation to thereby enable classical verification. For comparison, we follow up on the direct estimation of the $\eta$-state's energy. To this end, we implement the sub-circuit in the grey box from Fig.~\ref{Fig:ClassicalVerificationCircuit}a alongside additional operations on prover and clock qubit required to realize basis read-outs according to $Z_1Z_2$, $Z_1X_2$, $X_1Z_2$ and $X_1X_2$. Fig.~\ref{Fig:EtaStateEnergy} contains results on operator values (a) as well as energies covering both decision problems (b) and (c) considering the classical verification of $\alpha\rightarrow0$ (see main text) and $\alpha\rightarrow\pi/2$ (see Appendix~\ref{app:alphato1}) respectively. In both cases our experimental results undercut the threshold $1-b=0.4$ across the relevant region $\alpha\in[0,0.32]\cup[1.25,1.57]$, for which our decision problems hold. Comparably good results are obtained due to the less complex experiment using only one full-entangling $\mathsf{MS}$-gate plus on average eight single-qubit gates. Note that, these experiments were likewise performed on an eight-ion string using 400 experimental runs in each data point. Errors represent 1 standard-deviation from quantum projection noise.

We continue to estimate the $\eta$-state fidelity by incorporating error-rates inherent to the individual gates as previously done and thoroughly explained in the bottom part of Appendix~\ref{app:toolbox} - there considering the entire protocol. Here, the expected fidelity on the direct estimation reads:
\[
\mathcal{F} = \SI{0.998(1)}{}^8\cdot0.982(3) = 0.966(8).
\]
 We compare this number to results from Fig.~\ref{Fig:EtaStateEnergy}a by averaging $Z_1Z_2$ and $X_1X_2$ at $\alpha=\pi/2$ representing an estimate of the $\eta$-state's Bell-state fidelity. The result reads:
\[
\mathcal{F} \sim \frac{Z_1Z_2(\alpha=\pi/2)+X_1X_2(\alpha=\pi/2)}{2} = \frac{0.955(16)+0.935(17)}{2} = 0.945(12),
\]
and is again in good agreement with the above simplistic error-modelling.

\section{Noise model simulations}
\label{app:noisemodel}
This section aims to elaborate an error-model, which best describes the experimental data from the measurement rounds depicted in Fig.~\ref{Fig:MeasurementRounds} from the main text. Choosing a suitable error-model was done upon previous error-rate observations thoroughly discussed at the bottom of Appendix~\ref{app:toolbox}. Those observations distinctly reveal pairwise $\mathsf{MS}$-gates to limit the overall performance of our classical verification implementation. In contrast, single-qubit gates clearly make a smaller contribution, although having an approximately four times higher abundance in the final circuits. The analysis of individual $\mathsf{MS}$-gate pairs, depicted in Tab.~\ref{SITab:MSGateDecay}, discloses similar decay-rates in population and coherence - the latter characterizing the degree of loss in phase information. Hence, our findings support a simultaneous dephasing along X, Y and Z basis manifesting a so-called depolarizing channel. A fully depolarized state leads to a completely mixed state, which in the single qubit case is illustrated by shrinking the Bloch-sphere towards its center. Based on this, we worked out the following eight-qubit $\Gamma_\lambda=\Delta_\lambda^{\otimes 8}$ depolarizing channel to describe our classical verification results:
\begin{equation}\label{eq:depolchannel}
\begin{aligned}
\Gamma_\lambda(\rho_{\rm ideal})&=\Delta_\lambda^{\otimes 8}(\rho_{\rm ideal})\\
\Delta_\lambda(\sigma)&=\sum_{l=0}^3 K_l\sigma K_l^\dagger
\end{aligned}
\end{equation}
where $\Delta_\lambda$ are single-qubit depolarizing channels with Kraus operators
\[
\begin{aligned}
K_0&=\sqrt{1-3\lambda/4}\,{\bf 1}\,,\\
K_1&=\sqrt{\lambda/4}\,X\,,\\
K_2&=\sqrt{\lambda/4}\,Y\,,\\
K_3&=\sqrt{\lambda/4}\,Z\,.
\end{aligned}
\]

\begin{figure*}[ht]
    \centering
\includegraphics[width=0.9\textwidth]{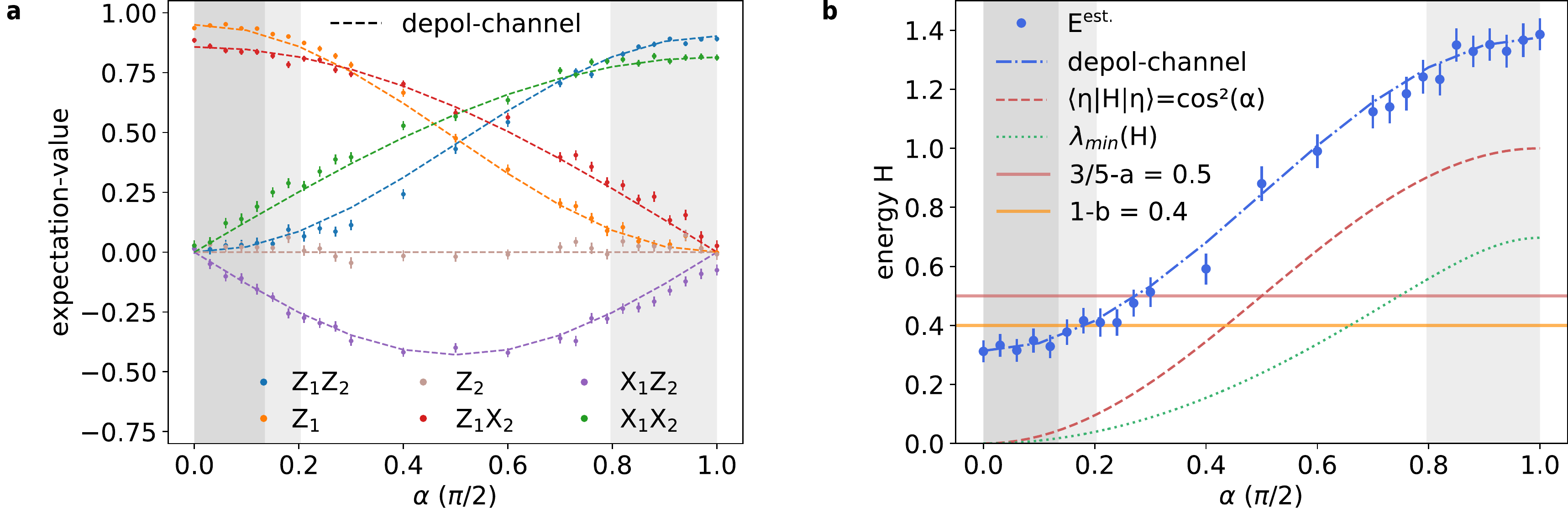}
    \caption{\textbf{Noise model simulation on the measurement rounds from Fig.~\ref{Fig:MeasurementRounds} in the main text.} We find good agreement in operator-values \textbf{a} as well as the $\eta$-state's total energy \textbf{b} between the experimental data and our noise-modelling represented by the depolarizing channel from Eq.~\eqref{eq:depolchannel} utilizing the best fitted parameter $\lambda=0.05$. In \textbf{b} we follow up on the numerical values of the ground-state energy $\lambda_\text{min}(\textsf{H}^{(\text{yes})})$ illustrated by the dotted-line according to $H=H_{\rm out}+6H_{\rm in}+3H_{\rm prop}$, see text for details.}
    \label{Fig:DepolNoise}
\end{figure*}

Hence $\lambda$ is the single-qubit depolarizing parameter. In Fig.~\ref{Fig:DepolNoise} we show the depolarizing channel using the best fitted rate given at $\lambda=0.05$ represented by dashed-lines on top of the data discussed in the main text. The good agreement between data and noise channel confirms our limitation to be depolarizing noise. According to the channel in Eq.~\eqref{eq:depolchannel} an ideal outcome in the measurement round with respect to the eight-qubit density matrix $\rho_\text{ideal}$ is expected at a probability of about $(1-\lambda)^8\approx0.66$. The total energy plot from (b) additionally depicts the ground-state energy of the implemented Hamiltonian $H=H_{\rm out}+6H_{\rm in}+3H_{\rm prop}$ with each term
given by~Eq.~\eqref{eq:H}. Note that, for this ground-state energies, i.e. the smallest numerical eigenvalues $\lambda_\text{min}(\textsf{H}^{(\text{yes})})$, the inequality $\lambda_\text{min}(H)>\bra\eta H\ket\eta-2/5$ holds across the entire $\alpha$ range, as discussed in Appendix~\ref{app:ineq}.

Moreover, this identical noise-model ($\lambda=0.05$) accurately images the experimental outcome on the extra decision problem presented in Appendix~\ref{app:alphato1}.

\bibliography{bibliography}

\end{document}